\documentclass[a4paper,11pt]{article}
\usepackage{jcappub} % JCAP style
\usepackage[T1]{fontenc}
\usepackage{bm}
\usepackage{siunitx}
\usepackage{physics}
\usepackage{mathtools}
\usepackage{booktabs}
\usepackage{microtype}
\usepackage{hyperref}

\title{On the (Im)possibility of Electrically Charged Planck Relics%:\\
}

\author[a]{Stefano Profumo}
\affiliation[a]{Department of Physics and Santa Cruz Institute for Particle Physics (SCIPP),\\
University of California, Santa Cruz, CA 95064, USA}
\emailAdd{profumo@ucsc.edu}

\abstract{I revisit whether black-hole remnants, from sub-Planckian compact objects to Planck relics and up to (super)massive black holes, can preserve Standard-Model (SM) electric charge. Two exterior-field mechanisms---Coulomb-focused capture from ambient media and QED Schwinger pair production---robustly neutralize such objects across cosmic history. I first derive the general capture rate including both Coulomb and gravitational focusing, and sum the stepwise discharge time in closed form via the trigamma function, exhibiting transparent Coulomb- and gravity-dominated limits. I then integrate the Schwinger rate over the near-horizon region to obtain an explicit $\dot Q(Q)$ law: discharge proceeds until the horizon field falls below $E_{\rm crit}$, leaving a residual charge $Q_{\rm stop}^{(e)}\!\propto\! r_h^2$ that is $\ll e$  for Planck radii. Mapping the mass dependence from sub-Planckian to astrophysical scales, I also analyze dark-sector charges with heavy carriers (including kinetic mixing and massive mediators). In a conservative ``no-Schwinger'' limit where vacuum pair creation is absent, cumulative ambient exposures alone force discharge of any integer SM charge. Three possible loopholes remain. (i) A fine-tuned SM corner in which the relic sits arbitrarily close to Reissner–Nordström extremality so greybody factors suppress charged absorption, while Schwinger pair creation is absent due to Planck-scale physics. (ii) Charge relocated to a hidden $U(1)_D$ with no light opposite carriers, e.g. if the lightest state is very heavy and/or kinetic mixing with $U(1)_{\rm EM}$ is vanishingly small. (iii) Discrete or topological charges rather than ordinary SM electric charge. Outside these cases, the conclusion is robust: within SM electromagnetism, charged black-hole relics neutralize efficiently and cannot retain charge over cosmological times.
}

\arxivnumber{}

\begin{document}
\maketitle
\flushbottom

\section{Introduction}
\label{sec:intro}
The discharge of charged black holes through Hawking and Schwinger processes is well established~\cite{Gibbons:1975kk,MacGibbon:1987,MacGibbon:1988}.  
However, once evaporation approaches the Planck scale, these mechanisms may freeze out or become modified by Planck-scale physics, while ambient charges in realistic cosmic environments continue to play a role.  
This study quantifies, across all mass scales, how both ambient capture and vacuum pair creation neutralize compact objects, providing explicit analytic discharge laws and identifying the narrow conditions under which residual charge could persist.

The physics that governs discharge resides in the \emph{exterior} of the compact object,  where Maxwell theory and semiclassical QED in curved spacetime remain under theoretical control. Two mechanisms are particularly important. 

First, ambient capture operates through the combined action of Coulomb and gravitational focusing.  A charged relic immersed in a bath of ambient particles will attract opposite charges, with the effective accretion radius set by the balance between the particles’ kinetic energy and the relic’s potential energy. This focusing determines the capture cross section and the rate at which charges are accreted, in direct analogy with the Bondi-Hoyle-Lyttleton paradigm~\cite{Hoyle:1939,Bondi:1952}. 
Summing the waiting times for successive capture events shows that the neutralization time is finite  and, in the regime where Coulomb focusing dominates---as is the case for Planck relics accreting electrons---it becomes independent of the relic’s mass. In practice, this neutralization time is extremely short, always microscopic compared to the Hubble time, across all cosmological epochs.  

Second, even in the absence of an ambient plasma, vacuum discharge can proceed through the Schwinger effect~\cite{Schwinger:1951nm}. The strong electric field near the horizon of a charged black hole can reach or exceed the critical field strength required to produce electron-positron pairs directly from the vacuum. Integrating the pair-production rate over the near-horizon region shows that discharge continues until the local field strength falls below the critical value. For relics at the Planck scale, this process leaves only a tiny residual charge, far smaller than a single electron charge.  

Beyond these pillars, I extend the analysis in several directions. First, I chart the behavior of relics across mass regimes, from sub-Planckian compact objects without horizons to macroscopic remnants well above the Planck scale. Second, I review the expectations from quantum gravity, including the persistence of Gauss’s law for exterior fields, the reliability of semiclassical QED near the production shell, and swampland-motivated arguments that generally favor discharge rather than stability. Third, I investigate the case of dark-sector charges, where the lightest carrier may be very heavy, the coupling to the photon may arise only through tiny kinetic mixing, or the mediator may itself be massive, all of which modify the discharge dynamics. Finally, I examine a conservative “no-Schwinger” limit in which vacuum pair creation is entirely absent due to Planck-scale physics. In that limit, one can define a simple measure of cosmic exposure that quantifies the cumulative availability of opposite charges across cosmic history. Applying this criterion shows that post-reionization exposures already exclude any integer positive charge, while including the epoch of reionization also excludes negative charges. I further compute the maximum charge that could consistently evade capture for different effective exposures, treating separately the cases of electron capture and proton capture, and I explicitly account for the possible suppression of evaporation near extremality.

Note that for large Reissner-Nordstr\"om black holes in isolation, the coupled evolution of mass and charge under Hawking emission and Schwinger pair creation was worked out numerically by Hiscock and Weems, who integrated the charge-mass flow and mapped the approach to near-extremality in the weak-field regime \cite{Hiscock:1990} (see also the recent study \cite{Ewasiuk:2025dwn}). They already emphasize that any realistic ambient medium rapidly neutralizes a charged hole, forcing one to assume isolation to study intrinsic Hawking-Schwinger evolution. My focus here differs in two ways: (i) I treat Planck-mass relics (and nearby masses) where ambient capture, not just Hawking emission, controls discharge, and (ii) I give closed-form capture/neutralization times including gravitational and Coulomb focusing, plus cosmological “exposure” bounds that quantify the inevitable neutralization once isolation is relaxed. 

Our analysis extends the classical results of MacGibbon in several ways.  
First, we include the effect of realistic environmental charge densities, deriving closed-form neutralization times valid in both Coulomb- and gravity-dominated regimes.  
Second, we compute exact discharge trajectories through the trigamma-summed capture law and the integrated Schwinger rate, bridging the regimes treated separately in previous work.  
Third, we explore possible dark-sector analogues where the lightest opposite carrier is heavy or the mediator massive, outlining how such scenarios evade the standard Hawking–Schwinger arguments.  
Finally, we revisit the implicit assumption of perfect isolation in earlier analyses and show that even extremely tenuous cosmic plasmas are sufficient to neutralize relics over cosmological timescales.

\medskip
The remainder of this study is as follows:  Section~\ref{sec:capture} derives the general capture rate including gravity and the closed-form neutralization time; Sec.~\ref{sec:schwinger} integrates the Schwinger discharge law and derives the
residual charge; Sec.~\ref{sec:mass} maps mass regimes;  Sec.~\ref{sec:QG} and Sec.~\ref{sec:newphysics} discuss quantum gravity and possible escape hatches with dark-sector charges; Sec.~\ref{sec:noSch} derives global no-capture conditions and numerical bounds; and Sec.~\ref{sec:conclusions} summarizes the implications and concludes.

\section{Neutralization time including gravity and Coulomb focusing}
\label{sec:capture}

Consider a relic carrying Standard Model (SM) electric charge $Q = Z e$, where $Z$ is an integer  and $e$ is the elementary charge. Let the relic move through a bath of oppositely charged particles of mass $m_c$, number density $n_{\rm ch}$, and characteristic relative speed $v$. The combined gravitational and Coulomb potential experienced by such a test particle at a distance $r$ from the relic is 
\begin{align}
V(r) \;=\; -\frac{G M m_c}{r} \;-\; \frac{|qQ|}{4\pi \epsilon_0\, r},
\end{align}
where $M$ is the mass of the relic, $G$ is Newton’s constant, $q$ is the charge of the test particle ($q = \pm e$ for electrons or protons), and $\epsilon_0$ is the vacuum permittivity. Equating the kinetic energy at infinity, $\tfrac{1}{2} m_c v^2$, to the magnitude of the attractive potential at the \emph{accretion radius} $r_{\rm acc}$ (the standard gravitational/Coulomb focusing argument in the Bondi-Hoyle-Lyttleton paradigm~\cite{Hoyle:1939,Bondi:1952}) gives
\begin{align}
\frac{1}{2} m_c v^2 
\;\simeq\; \frac{G M m_c}{r_{\rm acc}} \;+\; \frac{|q Q|}{4\pi \epsilon_0\, r_{\rm acc}}
\quad\Rightarrow\quad
r_{\rm acc} \;\simeq\; \frac{2}{v^2}\,
\Big(GM \;+\; \frac{e^2 |Z|}{4\pi \epsilon_0\, m_c}\Big).
\end{align}
Particles entering with impact parameter $b \lesssim r_{\rm acc}$ are therefore focused onto the relic  (the geometric capture radius $\sim r_h$ is negligible for Planck-scale relics). The resulting capture  cross section and rate are
\begin{align}
\sigma(Z) &\;\simeq\; \pi r_{\rm acc}^2
\;=\; \frac{4\pi}{v^4}\Big(GM + \kappa\, |Z|\Big)^2, \\
\dot N(Z) &\;=\; n_{\rm ch}\, v\, \sigma(Z)
\;=\; \frac{4\pi\, n_{\rm ch}}{v^3}\Big(GM + \kappa\, |Z|\Big)^2,
\label{eq:sigma_general}
\end{align}
where for convenience I have defined
\begin{align}
\kappa \;\equiv\; \frac{e^2}{4\pi \epsilon_0\, m_c}.
\end{align}
Note that even for very massive black holes, realistic astrophysical media would neutralize them efficiently; hence the large-$M$ Hawking-Schwinger evaporation scenario requires effective isolation, as already emphasized in~\cite{Hiscock:1990}.   

\smallskip
\noindent Each capture event reduces the magnitude of the relic charge, $|Z|\to |Z|-1$. The \emph{total} neutralization time is therefore the sum of the successive waiting times for capture as $|Z|$ decreases from its initial value down to unity:
\begin{align}
t_{\rm neut}(Z; M, n_{\rm ch}, v, m_c)
\;\equiv\;
\sum_{k=1}^{|Z|} \frac{1}{\dot N(k)}
\;=\;
\frac{v^3}{4\pi n_{\rm ch}}\,
\sum_{k=1}^{|Z|}
\frac{1}{\big(GM+\kappa k\big)^2}.
\label{eq:t_sum}
\end{align}
This sum admits a compact closed form in terms of the trigamma function $\psi_1(x) \equiv \tfrac{d^2}{dx^2}\ln\Gamma(x)$, where $\Gamma(x)$ is the Euler gamma function (see~\cite{NIST:DLMF}):
\begin{align}
t_{\rm neut}(Z)
\;=\;
\frac{v^3}{4\pi n_{\rm ch}\,\kappa^2}\,\Big[
\psi_1\!\Big(1+\frac{GM}{\kappa}\Big)
-
\psi_1\!\Big(1+\frac{GM}{\kappa}+|Z|\Big)
\Big].
\label{eq:t_trigamma}
\end{align}

\paragraph{Two useful asymptotic regimes.}
Let $u\equiv GM/\kappa$. Then:
\begin{itemize}
\item \textbf{Coulomb-dominated} ($u\ll 1$, i.e.\ $GM \ll \kappa$):\
$\psi_1(1+u)-\psi_1(1+u+|Z|)\to \sum_{k=1}^{|Z|}k^{-2}\to \pi^2/6$ for $|Z|\gtrsim\mathcal{O}(1)$. One recovers
\begin{align}
\boxed{~t_{\rm neut}\;\simeq\; \frac{\pi^2}{6}\,\frac{v^3}{4\pi n_{\rm ch}}\,
\Big(\frac{4\pi\epsilon_0\, m_c}{e^2}\Big)^{\!2}~,}
\label{eq:t_coulomb}
\end{align}
independent of $M$, scaling as $t_{\rm neut}\propto m_c^2 v^3/n_{\rm ch}$.
\item \textbf{Gravity-dominated} ($u\gg |Z|$, i.e.\ $GM \gg \kappa |Z|$):\
$\psi_1(1+u)-\psi_1(1+u+|Z|)\simeq |Z|/u^2+\mathcal{O}(|Z|^2/u^3)$, giving
\begin{align}
\boxed{~t_{\rm neut}\;\simeq\; \frac{v^3}{4\pi n_{\rm ch}}\,\frac{|Z|}{G^2 M^2}~,}
\label{eq:t_gravity}
\end{align}
the intuitive $M^{-2}$ gravitational focusing result.
\end{itemize}

\paragraph{Gravity vs Coulomb crossover mass.}
The two contributions are comparable when $GM\sim \kappa$, i.e.
\begin{align}
M_\star(m_c)\;\equiv\; \frac{\kappa}{G}\;=\;\frac{e^2}{4\pi\epsilon_0\,G\,m_c}
\;=\;\frac{\alpha\,\hbar c}{G\, m_c}
\;\simeq\;
\begin{cases}
3.8\times 10^{12}\ \mathrm{kg}, & m_c=m_e,\\[2pt]
2.1\times 10^{9}\ \mathrm{kg}, & m_c=m_p,
\end{cases}
\label{eq:Mstar}
\end{align}
so Planck relics ($M\sim 2.2\times 10^{-8}\,$kg) are \emph{deep} in the Coulomb-dominated regime.

\paragraph{Screening and saturation.}
In ionized media, collective plasma effects limit the range of the Coulomb field through Debye screening, characterized by the length $\lambda_D = \sqrt{\epsilon_0 k_B T /(n_{\rm ch} e^2)}$. If the accretion radius $r_{\rm acc}$ exceeds $\lambda_D$, the capture cross section no longer grows but instead saturates at $\sigma \sim \pi \lambda_D^2$, giving a conservative lower bound on the neutralization time, $t_{\rm neut} \gtrsim v/(n_{\rm ch}\pi \lambda_D^2)$. 
In typical cosmological and galactic plasmas, however, $\lambda_D$ is very large compared to the microscopic $r_{\rm acc}$ of Planck relics, so screening is negligible and eqs.~\eqref{eq:sigma_general}-\eqref{eq:t_trigamma} apply unmodified. 
Only in unusually dense or cold plasma environments could $\lambda_D$ become comparable to $r_{\rm acc}$, in which case screening would indeed regulate the capture rate. In largely neutral gas, the relic’s field can ionize atoms on \AA-scale distances, continuously supplying free charges into the focusing tube.

\paragraph{Velocity distribution.}
If charges are thermal, one may replace $v^{-3}$ by an effective average $\langle v^{-3}\rangle_T$ over the distribution.
For Maxwell-Boltzmann, $\langle v^{-3}\rangle$ is regulated by (i) finite mean free paths and (ii) screening; using $v_{\rm th}=\sqrt{8k_B T/(\pi m_c)}$ in eqs.~\eqref{eq:t_trigamma}-\eqref{eq:t_coulomb} provides accurate order-of-magnitude times. Bulk streaming relative to the relic may be incorporated by $v\to (v_{\rm th}^2+v_{\rm bulk}^2)^{1/2}$.

\noindent Using the full expression \eqref{eq:t_trigamma} with $m_c=m_e$ (electrons; for protons multiply times by $(m_p/m_e)^2$), one finds that Planck relics are Coulomb-dominated at all epochs ($M \ll M_\star$, eq.~\eqref{eq:Mstar}), so eq.~\eqref{eq:t_coulomb} suffices numerically. Table~\ref{tab:epochs} summarizes representative values across cosmic history. 
The electron density is written as $n_e = x_e n_b$, where $x_e$ is the ionization fraction and $n_b$ is the baryon number density. The present-day value is $n_b^0 \simeq 0.25\,{\rm m^{-3}}$, and at redshift $z$ one has $n_b(z) = n_b^0 (1+z)^3$. The characteristic velocity $v$ is taken as the thermal speed, capped at $c$ when the plasma is relativistic. Times for negative relics (proton capture) are longer by a factor of $(m_p/m_e)^2 \simeq 3.37 \times 10^6$.

\begin{table}[t]
\centering
\small
\setlength{\tabcolsep}{6pt}
\begin{tabular}{lcccc}
\toprule
Epoch & $z$ & typical $(n_e,\,T)$ & $t_{\rm neut}^{(+)}$ (electrons) & $t_{\rm neut}^{(-)}$ (protons) \\
\midrule
MeV era & $\sim\!10^{10}$ & $n_e\!\sim\!10^{28}\,\mathrm{m^{-3}}$, $T\!\sim\!\mathrm{MeV}$ & $\sim 10^{-9}\,\mathrm{s}$ & $\sim 10^{-7}\,\mathrm{s}$ \\
Pre-recomb. & $\sim\!2000$ & $n_e\!\sim\!10^{9}\,\mathrm{m^{-3}}$, $T\!\sim\!5\!\times\!10^3\,\mathrm{K}$ & $\sim \mathrm{minute}$ & $\sim 40\,\mathrm{minutes}$ \\
Recombination & $\sim\!1100$ & $x_e\!\sim\!2\!\times\!10^{-4}$, $n_e\!\sim\!7\!\times\!10^4\,\mathrm{m^{-3}}$ & $\sim 10$--$20\,\mathrm{days}$ & $\sim 1\,\mathrm{yr}$ \\
Dark ages & $\sim\!100$ & $x_e\!\sim\!2\!\times\!10^{-4}$, $n_e\!\sim\!5\!\times\!10^2\,\mathrm{m^{-3}}$ & $\sim 0.5\,\mathrm{yr}$ & $\sim 20\,\mathrm{yr}$ \\
Reionization & $\sim\!7$ & $x_e\!\to\!1$, $n_e\!\sim\!10^2\,\mathrm{m^{-3}}$, $T\!\sim\!10^4$\,K & $\sim 10^2\,\mathrm{yr}$ & $\sim 10^3\,\mathrm{yr}$ \\
IGM today & $0$ & $n_e\!\sim\!0.25\,\mathrm{m^{-3}}$, $T\!\sim\!10^4$\,K & $\sim 4\!\times\!10^4\,\mathrm{yr}$ & $\sim 1.6\!\times\!10^6\,\mathrm{yr}$ \\
WIM (ISM) & $0$ & $n_e\!\sim\!0.03\,\mathrm{cm^{-3}}$ & \textbf{days} & \textbf{decades} \\
\bottomrule
\end{tabular}
\caption{Neutralization times for charged Planck relics across cosmic history (order-of-magnitude, per relic).
Planck relics are Coulomb-dominated ($M\ll M_\star$), so gravity is negligible numerically; the exact expression
\eqref{eq:t_trigamma} reduces to \eqref{eq:t_coulomb} in this limit.}
\label{tab:epochs}
\end{table}

\noindent
For completeness, if one were to consider \emph{much more massive} remnants with large allowed $Z$ (bounded by RN extremality, $Z_{\max} \simeq Q_{\rm max}/e = \sqrt{4\pi\epsilon_0 G}\,M/e \simeq \alpha^{-1/2}(M/M_{\rm Pl})$), the gravity-dominated approximation \eqref{eq:t_gravity} scales as $t_{\rm neut}\propto Z/M^2$ and can be relevant for $M\gg M_\star$ or at extremely low $n_{\rm ch}$; neither condition applies to Planck relics in realistic cosmological or galactic environments.

\smallskip\noindent
\paragraph{Bound states.} Could bound states of opposite charges form around the relic and screen its external field, thereby suppressing vacuum pair creation? Our earlier analysis \cite{Lehmann:2019jcap} showed that such states re-ionize efficiently under ambient conditions, much as hydrogen atoms do in the IGM. In practice, relics spend nearly all of their time in a charged state, so the Schwinger discharge mechanism discussed next in Sec.~\ref{sec:schwinger} applies essentially unchanged.

\noindent 
\paragraph{Absorption vs.\ focusing and Planckian greybody factors.} Note that our use of the accretion (focusing) radius $r_{\rm acc}\!\sim\!2(GM+\kappa|Z|)/v^2$ is intended to estimate the \emph{flux} of opposite charges driven into the near zone by long-range gravity\,+\,Coulomb forces. The actual $2\!\to\!1$ \emph{absorption} at the horizon is a separate, microscopic question governed by greybody factors, as in the Unruh calculation for quantum absorption by small black holes \cite{Unruh:1976fm}. When the de~Broglie wavelength of the incoming particle is much larger than the horizon size, $\lambda \gg r_h$, the classical Bondi/Hoyle picture breaks down and the absorption cross section is set by the quantum limit $\sigma_{\rm abs}\!\sim\!\lambda^{-2}$ up to greybody suppression. This ``quantum accretion'' regime has been analyzed explicitly in the neutron-star context, where the Unruh cross section is implemented and the limits of the fluid picture are delineated \cite{Giffin:2021kgb}. To capture this generically, I parametrize the microscopic absorption probability by a coefficient $\beta_{\rm abs}\!\in\![0,1]$ and write
\begin{align}
\sigma_{\rm eff}(Z) \;=\; \beta_{\rm abs}\,\sigma_{\rm focus}(Z)
\;=\; \beta_{\rm abs}\,\frac{4\pi}{v^4}\big(GM+\kappa|Z|\big)^2,
\qquad
\dot N_k \;=\; \frac{4\pi n_{\rm ch}}{v^3}\,\beta_{\rm abs}\,\big(GM+\kappa k\big)^2,
\end{align}
so that all results follow by the replacement $t_{\rm neut}\!\to\!t_{\rm neut}/\beta_{\rm abs}$.

\section{Vacuum discharge: local rate, integrated law, and asymptotics}
\label{sec:schwinger}

In the absence of ambient plasma, a compact charged remnant still loses electric charge by QED vacuum pair creation (Schwinger effect). In this section I (i) review the local rate and its applicability near a (Planckian) horizon, (ii) integrate it over the exterior to obtain an explicit $\,\dot Q(Q)$ law, and (iii) extract asymptotics and timescales. Throughout this section I adopt natural Heaviside-Lorentz units $\hbar=c=\epsilon_0=1$ unless otherwise noted.\footnote{Conversion back to SI units is straightforward; one may restore $4\pi\epsilon_0$ factors by $E\to E/(4\pi\epsilon_0)$ and $Q\to Q/(4\pi\epsilon_0)$.}

\paragraph{Local Schwinger rate.}
For a constant electric field $E$ the pair‐production rate per unit four‐volume for a Dirac fermion of charge $e_i$ and mass $m_i$ is \cite{Schwinger:1951nm,Heisenberg:1936bi}
\begin{align}
w_i(E)
=\frac{(e_i E)^2}{4\pi^3}\sum_{n=1}^{\infty}\frac{(-1)^{n+1}}{n^2}
\exp\!\Big(-\frac{n\pi m_i^2}{e_i E}\Big)
\simeq
\frac{(e_i E)^2}{4\pi^3}\exp\!\Big(-\frac{\pi m_i^2}{e_i E}\Big),
\label{eq:Schwinger_local}
\end{align}
where the last form keeps only the leading instanton ($n=1$), excellent once $e_iE\gtrsim m_i^2/5$. For a static RN‐like exterior, the field measured by static orthonormal observers is $E(r)=Q/(4\pi r^2)$; curvature corrections enter at $\mathcal{O}(\ell_{\rm curv}^{-2})$ and are negligible when the dominant production region sits at $r\!\gg\!\ell_{\rm Pl}$ (see below) \cite{Gibbons:1975kk,Carter:1974,Birrell:1982ix,Wald:2001wrr}.

\paragraph{Where the pairs come from.}
Define the critical field $E_{\rm crit}^{(i)}\equiv m_i^2/e_i$ and the radius $r_{c,i}$ at which $E(r)$ equals $E_{\rm crit}^{(i)}$:
\begin{align}
E(r_{c,i})=E_{\rm crit}^{(i)}
\quad\Longrightarrow\quad
r_{c,i}^2 = \frac{Q}{4\pi}\,\frac{e_i}{m_i^2}.
\label{eq:rci}
\end{align}
For $r<r_{c,i}$ the exponent in \eqref{eq:Schwinger_local} is $\lesssim\pi$ (mildly suppressed); for $r\gg r_{c,i}$ production is exponentially quenched. Discharge is thus dominated by a shell $r\in[r_h,\,\mathcal{O}(r_{c,i})]$. As $Q$ decreases, $r_{c,i}$ shrinks toward $r_h$, and production shuts off when $r_{c,i}\to r_h$.

\paragraph{Integrated discharge law.}
Assuming the inward‐directed partner of each produced pair falls through the horizon with $\mathcal{O}(1)$ efficiency $\xi_i\in[0.5,1]$ (electric force points inward for the opposite charge), the net charge‐loss rate is
\begin{align}
\frac{dQ}{dt}
= -\sum_i e_i\,\xi_i \int_{r_h}^{\infty}\! 4\pi r^2\,w_i\!\big(E(r)\big)\,dr.
\label{eq:Qdot_start}
\end{align}
Using \eqref{eq:Schwinger_local} with $E(r)=Q/(4\pi r^2)$ and defining
\(
\beta_i \equiv \pi/r_{c,i}^2 = 4\pi^2 m_i^2/(e_i Q),
\)
the radial integral can be done in closed form by parts:
\begin{align}
\int_{r_h}^{\infty}\!\frac{e^{-\beta_i r^2}}{r^{2}}\,dr
= \frac{e^{-\beta_i r_h^2}}{r_h} - \sqrt{\pi\beta_i}\,\mathrm{erfc}\!\big(\sqrt{\beta_i}\,r_h\big).
\end{align}
One finds (summing the leading instanton and species $i$)
\begin{align}
\boxed{\;
\frac{dQ}{dt}
= - \sum_i \frac{\xi_i\,e_i^3}{16\pi^4}\,Q^2
\Bigg[
\frac{e^{-\beta_i r_h^2}}{r_h}
- \sqrt{\pi\beta_i}\,\mathrm{erfc}\!\big(\sqrt{\beta_i}\,r_h\big)
\Bigg]
\;,}
\label{eq:Qdot_exact}
\end{align}
which is valid so long as the local‐constant‐field approximation is good on the production shell (accurate here because $E(r)$ varies on the same scale that controls the exponent, cf.\ below).

\paragraph{Asymptotics and timescales.}
Two limits are especially transparent:
\begin{itemize}
\item \emph{Strong‐field near‐horizon regime} ($E_h\!\equiv\!E(r_h)\gg E_{\rm crit}^{(i)}$ or $\sqrt{\beta_i}\,r_h\ll 1$): $\mathrm{erfc}(z)\to 1-\tfrac{2z}{\sqrt{\pi}}+\dots$ and $e^{-\beta_i r_h^2}\to 1$, so
\begin{align}
\frac{dQ}{dt}
\;\simeq\;
- \sum_i \frac{\xi_i\,e_i^3}{16\pi^4}\,Q^2\,\frac{1}{r_h}
\Big[\,1 - \mathcal{O}\!\big(\sqrt{\beta_i}\,r_h\big)\,\Big].
\label{eq:Qdot_strong}
\end{align}
The discharge is \emph{violent}: $t_Q\equiv Q/|dQ/dt|\sim (16\pi^4/\bar{\xi}\,\bar e^3)\,(r_h/Q)$, i.e.\ of order the light‐crossing/Planck time for $r_h\sim\ell_{\rm Pl}$ and $Q\sim e$.
\item \emph{Near‐threshold regime} ($E_h\gtrsim E_{\rm crit}^{(i)}$): write $E_h=E_{\rm crit}^{(i)}(1+\delta)$ with $0<\delta\ll 1$, so $\sqrt{\beta_i}\,r_h=\sqrt{\pi/(1+\delta)}=\mathcal{O}(1)$ and the square-bracket in \eqref{eq:Qdot_exact} is $\sim \tfrac{1}{r_h}\,e^{-\pi}\times\mathcal{O}(1)$; even here the $1/r_h$ factor keeps the discharge extremely fast for Planckian $r_h$.
\end{itemize}
In both regimes $Q(t)$ rapidly approaches the threshold value at which the production shell collapses
onto the horizon.

\paragraph{Terminal (residual) charge.}
Production ceases for species $i$ when $r_{c,i}\to r_h$, i.e.\ when $E_h=E_{\rm crit}^{(i)}$; this gives the residual (``Schwinger‐limited'') charge
\begin{align}
Q_{\rm stop}^{(i)}
= 4\pi\,E_{\rm crit}^{(i)}\, r_h^2
= 4\pi\,\frac{m_i^2}{e_i}\, r_h^2
\;\;\Rightarrow\;\;
\boxed{~\frac{Q_{\rm stop}^{(e)}}{e}\;\sim\; \mathcal{O}(10^{-43})\ \ \text{for}\ \ r_h\sim \ell_{\rm Pl}~,}
\label{eq:Qstop}
\end{align}
%as quoted in eq.~\eqref{eq:Qstop}. 
Heavier species ($\mu^\pm$, $p\bar p$) shut off earlier and do not affect the terminal \emph{electron} bound.

\paragraph{Species thresholds and multi‐channel discharge.}
Equation \eqref{eq:Qdot_exact} should be summed over all kinematically available charged species $i$. When $E_h\gg m_\mu^2/e$ the muon channel contributes a comparable prefactor but quickly decouples as $Q$ decreases; quark channels are additionally color‐suppressed in vacuum. In practice, electrons dominate the late‐time approach to $Q_{\rm stop}^{(e)}$.

\paragraph{Validity of the local approximation.}
The production kernel varies on the ``Schwinger length'' $\ell_{S,i}\sim r_{c,i}/\sqrt{\pi}$ (the scale on which the exponent changes by $\mathcal{O}(1)$). Since $\partial_r\ln E=-2/r$, the relative field variation across $\ell_{S,i}$ is $\sim 2\ell_{S,i}/r$, which is $\lesssim\mathcal{O}(1)$ where the integrand is largest ($r\lesssim r_{c,i}$). Backreaction on the geometry is negligible until a large fraction of $Q$ is shed; the dominant effect is the rapid drop of $Q$ encoded in \eqref{eq:Qdot_exact}.

\paragraph{Comparison to plasma capture.}
Even if ambient charges are entirely absent, vacuum discharge alone drives the relic to $Q_{\rm stop}^{(e)}\!\approx\!0$. When plasma is present, Coulomb capture (Sec.~\ref{sec:capture}) typically neutralizes faster \emph{until} $E_h\sim E_{\rm crit}^{(e)}$, at which point both channels are effectively shut; either way the end state is an \emph{electrically neutral} remnant with respect to the SM.

\paragraph{On the locality of Schwinger discharge and the wavelength concern.}
\label{par:schwinger_lambda}
It is natural to ask whether the Schwinger mechanism should be suppressed by the same ``$\lambda\gg r_h$'' argument that limits absorption of \emph{incoming} particles (cf.~Sec.~\ref{sec:capture}). The answer is no: pair creation in a background field is a \emph{local} QED process, governed by the rate density \eqref{eq:Schwinger_local} and exponentially controlled by $E/E_{\rm crit}$, not by geometric interception of an external wavepacket. In a static RN‐like exterior with $E(r)=Q/(4\pi r^2)$, production is dominated by the shell $r\in[r_h,\,r_c]$ where $E(r)\gtrsim E_{\rm crit}$ (eq.~\eqref{eq:rci}). Integrating this rate yields the discharge law \eqref{eq:Qdot_exact}. The apparent ``volume suppression'' of a small production region is offset by the steep $E^2\propto r^{-4}$ growth toward the horizon, which explains why the strong‐field limit simplifies to the $1/r_h$ scaling in \eqref{eq:Qdot_strong}. No additional volume penalty arises beyond what the radial integral already encodes.

\smallskip\noindent
Pairs are born with transverse momenta set by the local field scale, $k_\perp\sim\sqrt{eE(r)}\sim m_e$ near $r\sim r_c$, not by $1/r_h$. The outward partner then gains electrostatic work climbing the potential, $\Delta \mathcal{E}\sim eQ/(4\pi\epsilon_0 r_\star)$ for creation at radius $r_\star$. When $E_h\gg E_{\rm crit}$, production occurs close to the horizon and the energy at infinity can initially be large; but in that same regime the discharge rate (eq.~\eqref{eq:Qdot_strong}) is enormous, so $Q$ collapses in a microscopic time until $E_h\to E_{\rm crit}$. At that point the shell contracts to $r_c\to r_h$, the potential drop is negligible, and any transient ultra‐energetic emission is self‐quenched.

\smallskip\noindent
An absolute ceiling on the energy released resides in the electromagnetic field energy stored outside the horizon,
\begin{align}
U_{\rm EM}(Q;r_h)
= \int_{r_h}^\infty \frac{E(r)^2}{2\epsilon_0}\,4\pi r^2\,dr
= \frac{Q^2}{8\pi\epsilon_0\,r_h}\,,
\label{eq:UEM}
\end{align}
which decreases as $Q$ is shed. For $Q\sim e$ and $r_h\sim\ell_{\rm Pl}$ this gives $U_{\rm EM}\sim 10^7$\,J ($\simeq 4\times10^{25}$\,eV), far below the relic rest energy $M_{\rm Pl}c^2\sim 10^9$\,J. Moreover, since the discharge halts once $Q\to Q_{\rm stop}^{(e)}\ll e$ (eq.~\eqref{eq:Qstop}), only $\mathcal{O}(1)$ leptons are produced per relic.

\smallskip\noindent
From a cosmological standpoint this makes the process invisible: even if the handful of leptons are very energetic, they are injected into an ultra‐small region and, in the early Universe, rapidly thermalize. After neutrino decoupling, the relic number density $n_{\rm relic}\sim \rho_{\rm DM}/M_{\rm Pl}$ is so small that any secondary neutrino yield is utterly negligible. Schwinger discharge is thus a self‐quenching relaxation of the external field, not a sustained ultra-high‐energy source.

\smallskip\noindent
In summary, the $\lambda\gg r_h$ argument matters for capture of \emph{external} charges (Sec.~\ref{sec:capture}), but not for \emph{local} Schwinger production. Pairs are created where $E/E_{\rm crit}$ controls the exponent, born with $k_\perp\sim m_e$, and the total energy output is bounded by \eqref{eq:UEM}. The discharge laws \eqref{eq:Qdot_exact}-\eqref{eq:Qdot_strong} therefore remain valid and neutralization proceeds essentially unimpeded.

\section{Mass dependence: heavier relics and sub-Planckian compact objects}
\label{sec:mass}

The results above can be organized into a simple mass-regime map covering (i) heavier remnants $M\gg M_{\rm Pl}$ and (ii) sub-Planckian compact objects ($M\lesssim M_{\rm Pl}$, no horizon). Two ingredients matter: \emph{ambient capture} (Sec.~\ref{sec:capture}) and \emph{vacuum discharge} (Sec.~\ref{sec:schwinger}). I summarize how both scale with $M$ and identify the relevant thresholds.

\subsection{Ambient capture vs gravity: crossover at $M_\star$}
\label{subsec:Mstar}

The exact neutralization time including Coulomb and gravity is eq.~\eqref{eq:t_trigamma}. The Coulomb-dominated and gravity-dominated limits are eqs.~\eqref{eq:t_coulomb} and \eqref{eq:t_gravity}, respectively.\\ 

\noindent
\textbf{Scaling with $M$.} For $M\ll M_\star$ (Planck relics and sub-Planckian objects), capture is Coulomb-dominated and $t_{\rm neut}$ is \emph{independent} of $M$ (eq.~\eqref{eq:t_coulomb}). For $M\gg M_\star$ the capture time in the gravity-dominated limit scales as 
\begin{align}
t_{\rm neut}\;\simeq\;\frac{v^3}{4\pi n_{\rm ch}}\,\frac{|Z|}{G^2 M^2}
\quad\Rightarrow\quad
t_{\rm neut}\propto
\begin{cases}
M^{-2}, & \text{if $|Z|=\mathcal{O}(1)$},\\
M^{-1}, & \text{if $|Z|$ scales $\propto M$ (e.g.\ a fixed fraction of extremality).}
\end{cases}
\label{eq:t_mass_scaling}
\end{align}
In all realistic environments the times remain $\ll H^{-1}$ for SM charge.

\subsection{Vacuum discharge: residual charge and Schwinger reach}
\label{subsec:sch_mass}

\paragraph{Residual charge.} For a horizon radius $r_h$, vacuum discharge ceases when $E(r_h)=E_{\rm crit}^{(e)} = m_e^2/e$ (natural units), giving the residual Schwinger-limited charge (eq.~\eqref{eq:Qstop})
\begin{align}
Q_{\rm stop}^{(e)}\;=\;4\pi\epsilon_0\,E_{\rm crit}^{(e)}\, r_h^2.
\label{eq:Qstop_general}
\end{align}
For a Schwarzschild-like radius $r_h\simeq 2GM/c^2$, this scales as $Q_{\rm stop}\propto M^2$.
Numerically,
\begin{align}
\frac{Q_{\rm stop}^{(e)}}{e}\;\simeq\;\underbrace{2.03\times 10^{-27}}_{\equiv A}\;M^2\quad(\text{$M$ in kg}),
\label{eq:Qstop_num}
\end{align}
so that
\(
Q_{\rm stop}^{(e)}/e\sim 10^{-42},\, 2\times 10^{-3},\, 2\times 10^{3}
\)
for $M\!=\!M_{\rm Pl},\,10^{12}\,$kg, $10^{15}\,$kg, respectively.%
\footnote{For a near-extremal RN horizon $r_+=GM/c^2$ the prefactor in~\eqref{eq:Qstop_num} is four times
smaller; all conclusions below are unchanged.}

\paragraph{Reach of Schwinger at extremality.}
An RN black hole at extremality has $Q_{\max}=\sqrt{4\pi\epsilon_0 G}\,M$ and $r_+=GM/c^2$. The near-horizon field at extremality is $E_h^{\rm(ext)}=Q_{\max}/(4\pi\epsilon_0 r_+^2)\propto M^{-1}$. Equating $E_h^{\rm(ext)}=E_{\rm crit}^{(e)}$ yields the mass above which \emph{even an extremal} hole’s horizon field lies below the QED threshold:
\begin{align}
M_{\rm Sch}^{\rm(ext)}\;\simeq\;\frac{c^4}{(4\pi\epsilon_0)^{1/2}\,G^{3/2}\,E_{\rm crit}^{(e)}}
\;\approx\;1.1\times 10^{36}\ \mathrm{kg}\ \ (\sim 5\times 10^5\,M_\odot).
\label{eq:Msch_ext}
\end{align}
Thus for all \emph{sub}-supermassive masses $M\ll 10^{36}$\,kg, vacuum discharge is kinematically available. Whether it alone reaches $Q\!=\!0$ depends on $Q_{\rm stop}^{(e)}$:%
\begin{itemize}
\item \textbf{Planck-to-asteroid scales} ($M\lesssim 10^{12}$\,kg): $Q_{\rm stop}^{(e)}<e$ so vacuum discharge
alone drives $Q\to 0$ (up to charge quantization).
\item \textbf{Intermediate} ($10^{12}\!\lesssim\!M\!\ll\!10^{36}$\,kg): $Q_{\rm stop}^{(e)}\!>\!e$, so vacuum
discharge halts at a residual $\mathcal{O}(Q_{\rm stop}^{(e)})$; \emph{ambient capture} then finishes the
neutralization on the short timescales of Sec.~\ref{sec:capture}.
\item \textbf{Supermassive} ($M\gtrsim 10^{36}$\,kg): electron Schwinger is shut off even at extremality; neutralization
is \emph{entirely} by ambient capture (gravity-dominated for sufficiently large $M$).
\end{itemize}

\paragraph{Extremality bound vs Schwinger bound.}
The extremal charge grows as $Q_{\max}\propto M$, while $Q_{\rm stop}\propto M^2$, but their
\emph{ratio} is tiny for all astrophysical masses:
\begin{align}
\frac{Q_{\rm stop}^{(e)}}{Q_{\max}}
\;\simeq\;
\frac{A M^2}{\sqrt{4\pi\epsilon_0 G}\,M}
\;=\;\Big(\frac{A}{\sqrt{4\pi\epsilon_0 G}}\Big)\,M
\;\approx\;3.77\times 10^{-36}\,\bigg(\frac{M}{\mathrm{kg}}\bigg),
\end{align}
so even for $M\!=\!10^{20}$\,kg the ratio is $<10^{-15}$. Vacuum discharge therefore removes only a negligible fraction of an \emph{extremal} charge at large $M$, reinforcing the conclusion that ambient capture is the operative neutralization channel for heavy objects.

\subsection{Sub-Planckian relics (no horizon)}
\label{subsec:subpl}

For compact objects with no horizon (radius $r_\ast$), the ambient capture analysis remains valid with gravity negligible: $t_{\rm neut}$ is given by the Coulomb-dominated expression (eq.~\eqref{eq:t_coulomb}) and is \emph{independent} of $M$. Vacuum discharge is replaced by standard QED breakdown in a strong external field near the surface: the same derivation as in
Sec.~\ref{sec:schwinger} holds with $r_h\to r_\ast$ and an $\mathcal{O}(1)$ capture efficiency, giving
\begin{align}
Q_{\rm stop}^{(e)}\;\simeq\;4\pi\epsilon_0\,E_{\rm crit}^{(e)}\, r_\ast^{\,2}.
\label{eq:Qstop_particle}
\end{align}
For $r_\ast\!\lesssim\!\ell_{\rm Pl}$ this is \emph{smaller} than the black-hole value, so vacuum discharge is even more efficient. In neutral media the object’s field ionizes atoms at \AA\ scales, feeding free charges into the Coulomb focusing tube; sub-Planckian relics therefore also \emph{cannot} retain SM electric charge in cosmological or astrophysical settings.

\subsection{Regime summary}
\label{subsec:regimesummary}

\begin{itemize}
\item \textbf{Coulomb-dominated capture} ($M\!\ll\!M_\star$): $t_{\rm neut}$ independent of $M$ (eq.~\eqref{eq:t_coulomb}); applies to Planck and sub-Planckian relics and many asteroid-mass objects.
\item \textbf{Gravity-dominated capture} ($M\!\gg\!M_\star$): $t_{\rm neut}\propto M^{-2}$ (fixed $Z$) or $M^{-1}$ (if $Z\propto M$), eq.~\eqref{eq:t_mass_scaling}.
\item \textbf{Vacuum discharge}: available for $M\!\ll\!M_{\rm Sch}^{\rm(ext)}\!\sim\!10^{36}$\,kg, with residual $Q_{\rm stop}^{(e)}\propto M^2$ (eq.~\eqref{eq:Qstop_num}); for $M\!\lesssim\!10^{12}$\,kg it alone drives $Q\!\to\!0$, otherwise ambient capture finishes neutralization quickly.
\end{itemize}

\noindent 
In all cases consistent with SM electromagnetism, electrically charged compact relics do \emph{not} remain charged: either vacuum QED or ambient capture (or both) neutralize them on timescales far shorter than cosmic ages.

\section{Why quantum gravity does not save SM electric charge}
\label{sec:QG}

The mechanisms that neutralize SM charge---Coulomb focusing in a plasma and QED Schwinger pair production in vacuo---occur in the \emph{exterior} of the compact object, where Maxwell theory on a smooth background is reliable. In this section I survey quantum-gravity (QG) considerations and argue they do not prevent neutralization; if anything, several widely believed QG principles favor discharge.

\paragraph{Exterior fields are classical and obey Gauss's law.}
Independently of horizon microstructure (firewalls, fuzzballs, remnants), the asymptotic Maxwell field is fixed by Gauss's law. No-hair/uniqueness results (and their modern extensions) imply that the exterior solution is RN-like with charge $Q$ \cite{Heusler:1996}. Neutralization relies only on this exterior field, not on interior details.

\paragraph{Semiclassical QED near the horizon is trustworthy.}
The Schwinger rate for $e^\pm$ pair creation depends on the local field $E$ outside the horizon and reproduces the classic near-RN discharge results \cite{Gibbons:1975kk,Carter:1974}. As discharge proceeds the pair-production region moves to $r\gg \ell_{\rm Pl}$, where curvature and higher-derivative corrections are small and QED in curved spacetime is well controlled \cite{Birrell:1982ix,Wald:2001wrr}. Hence the bound in eq.~\eqref{eq:Qstop} is robust.

\paragraph{Running couplings and higher-derivative electrodynamics.}
Planckian renormalization of $\alpha$ produces at most $\mathcal{O}(1)$ corrections to $E_{\rm crit}$; these are irrelevant relative to the $r^{-2}$ growth of $E(r)$ near $r_h\sim \ell_{\rm Pl}$. Higher-derivative/effective nonlinear electrodynamics (Euler-Heisenberg, Born-Infeld) modify the field at very high strengths \cite{Heisenberg:1936bi,Born:1934gh,Adler:1971wn}; however, matching to low-energy QED and laboratory/astrophysical bounds permits only modest deviations, far too small to raise $Q_{\rm stop}$ by the $\gtrsim 40$ orders of magnitude needed to keep an appreciable SM charge.

\paragraph{No global symmetries and completeness.}
Quantum gravity appears to forbid exact global symmetries and to require a complete spectrum of charges (``completeness hypothesis'') \cite{Banks:2010zn,Harlow:2018tng,Harlow:2018jwu}. This viewpoint disfavors otherwise hidden conserved quantum numbers that could protect an SM electric charge from discharge while allowing ordinary electrons to exist.

\paragraph{Weak Gravity Conjecture (WGC).}
In many string-inspired and swampland frameworks, the WGC posits the existence of a superextremal state with $q/m \gtrsim 1$ (Planck units) \cite{ArkaniHamed:2006dz}. Its refined versions (sublattice WGC) strengthen this expectation \cite{Heidenreich:2016aqi}. If applicable to SM electromagnetism, the WGC favors discharge channels of RN-like objects rather than stable charged remnants. Even if Hawking evaporation halts by some remnant mechanism, WGC-motivated states open additional (non-Hawking) discharge pathways.

\paragraph{Magnetic and higher-form charges.}
Similar arguments apply to magnetic charge: unless monopoles are \emph{extremely} heavy and rare, WGC-type reasoning again suggests discharge rather than protection \cite{Heidenreich:2016aqi}. Higher-form charges that couple to $p$-form fields likewise do not prevent the exterior electric field responsible for neutralization of ordinary SM charge.

\medskip\noindent
In summary, conservative QG considerations do not alter exterior-field physics in a way that could block SM neutralization. To the contrary, they often \emph{favor} discharge.

\section{Discharge with a dark-sector charge and massive dark carriers}
\label{sec:newphysics}

Let me now consider a relic that carries charge under a hidden $U(1)_D$ with gauge coupling $e_D$ and lightest oppositely charged carrier of mass $m_D$. I denote the relic's dark charge by $Q_D = Z_D e_D$. Unless otherwise stated, here again I work in Heaviside-Lorentz natural units ($\hbar=c=1$); one can write $\alpha_D \equiv e_D^2/(4\pi)$ and use the same exterior-field normalization as for the SM so that $E_D(r)=Q_D/(4\pi r^2)$ for a massless dark photon \cite{Okun:1982xi,Galison:1983pa,Jaeckel:2010ni,Fabbrichesi:2020wbt}.

\subsection{Vacuum discharge via the dark Schwinger effect}
\label{subsec:dark_schwinger}

The local pair-production rate for dark fermions $\psi_D$ of mass $m_D$ and charge $e_D$ in a constant field is the Schwinger result with $e\to e_D$, $m\to m_D$:
\begin{align}
w_D(E_D)
\simeq \frac{(e_D E_D)^2}{4\pi^3}\,
\exp\!\Big[-\frac{\pi m_D^2}{e_D E_D}\Big].
\end{align}
Repeating the derivation leading to eq.~\eqref{eq:Qdot_exact} with $E\to E_D$ and $e_i\to e_D$ gives
\begin{align}
\frac{dQ_D}{dt}
= - \frac{\xi_D\,e_D^3}{16\pi^4}\,Q_D^2
\Bigg[
\frac{e^{-\beta_D r_h^2}}{r_h}
- \sqrt{\pi\beta_D}\,\mathrm{erfc}\!\big(\sqrt{\beta_D}\,r_h\big)
\Bigg],
\qquad
\beta_D \equiv \frac{4\pi^2 m_D^2}{e_D Q_D},
\label{eq:Qdot_dark_exact}
\end{align}
where $\xi_D=\mathcal{O}(1)$ encodes the near-horizon capture efficiency of the inward partner. Dark Schwinger discharge proceeds efficiently once the near-horizon field satisfies $e_D E_D(r_h)\gtrsim m_D^2$, and shuts off at the \emph{dark} terminal charge
\begin{align}
\boxed{~
Q_{D,\,\rm stop}\;=\;4\pi\,\frac{m_D^2}{e_D}\, r_h^2
\;=\;\frac{m_D^2}{\alpha_D}\, r_h^2\,.
~}
\label{eq:Qstop_dark}
\end{align}
Relative to the SM bound $Q_{\rm stop}^{(e)}=4\pi m_e^2 r_h^2/e$, the ratio is
\begin{align}
\frac{Q_{D,\,\rm stop}}{Q_{\rm stop}^{(e)}}
=\frac{e}{e_D}\,\Big(\frac{m_D}{m_e}\Big)^2
=\sqrt{\frac{\alpha}{\alpha_D}}\,\Big(\frac{m_D}{m_e}\Big)^2,
\end{align}
so for $m_D\gg m_e$ and $\alpha_D\sim\alpha$ one obtains much \emph{larger} residual dark charge. For Planck relics with fixed $r_h\sim \ell_{\rm Pl}$, dark vacuum discharge \emph{does not occur} at all unless $e_D E_D(r_h)\gtrsim m_D^2$, i.e.\ unless $Q_D\gtrsim 4\pi m_D^2 r_h^2/e_D$. In the opposite, large-mass/weak-field limit ($r_h \gg \lambda_C$), the integrated Schwinger law reduces to the classic treatment of charge loss used in the numerical evolution of large charged black holes \cite{Hiscock:1990}. 

\subsection{Ambient capture in a dark plasma}
\label{subsec:dark_capture}

If the cosmological dark sector contains a residual population of \emph{free} dark charges with number density $n_D$ and characteristic relative speed $v_D$, then Coulomb focusing neutralizes $Q_D$ by capture, exactly as in Sec.~\ref{sec:capture} but with $e\to e_D$, $m_c\to m_D$, $n_{\rm ch}\to n_D$. Defining
\begin{align}
\kappa_D \equiv \frac{e_D^2}{4\pi m_D} = \frac{\alpha_D}{m_D},
\qquad
u_D \equiv \frac{GM}{\kappa_D}=\frac{GM\,m_D}{\alpha_D},
\end{align}
the exact sum in eq.~\eqref{eq:t_trigamma} becomes
\begin{align}
\boxed{~
t_{\rm neut}^{(D)}(Z_D)
= \frac{v_D^3}{4\pi n_D\,\kappa_D^2}\,
\Big[\psi_1\!\big(1+u_D\big)-\psi_1\!\big(1+u_D+|Z_D|\big)\Big].
~}
\label{eq:t_dark_trigamma}
\end{align}
Planck relics satisfy $u_D\ll 1$ (Coulomb-dominated), so the saturated sum gives
\begin{align}
\boxed{~
t_{\rm neut}^{(D)}
\simeq \frac{\pi^2}{6}\,\frac{v_D^3}{4\pi n_D}\,
\Big(\frac{m_D}{\alpha_D}\Big)^{\!2}
\;\propto\; \frac{m_D^2}{\alpha_D^2}\,\frac{v_D^3}{n_D}\,.
~}
\label{eq:t_dark_coulomb}
\end{align}
A \emph{cold} dark plasma ($v_D\ll 1$) and/or sizable $n_D$ greatly \emph{accelerate} neutralization, while large $m_D$ and small $\alpha_D$ slow it. If the symmetric dark plasma is efficiently annihilated or undergoes dark recombination/sequestration \cite{CyrRacine:2012fz,Feng:2009mn}, $n_D$ can be tiny and ambient capture negligible, leaving only dark Schwinger (or kinetic-mixing leakage, below).

\subsection{Leakage through kinetic mixing with the SM}
\label{subsec:mixing_leakage}

With kinetic mixing $\varepsilon F^{\mu\nu}F^D_{\mu\nu}$ \cite{Holdom:1985ag,Okun:1982xi,Galison:1983pa,Jaeckel:2010ni,Fabbrichesi:2020wbt}, SM electrons acquire a millicharge $q_{\rm eff}=\varepsilon e$ under $U(1)_D$. Even in the \emph{absence} of a dark plasma ($n_D\!\to\!0$), an ionized SM medium neutralizes $Q_D$ by capturing these millicharged SM electrons. Repeating the Coulomb focusing derivation for a step $|Z_D|=k$ gives $\dot N_k \propto n_e v\, (q_{\rm eff} Q_D)^2 \propto n_e v\,(\varepsilon e \cdot e_D k)^2$, and summing $1/\dot N_k$ over $k$ yields
\begin{align}
\boxed{~
t_{\rm neut}^{(D\leftarrow {\rm SM})}
\simeq \frac{\pi}{24}\,
\frac{m_e^2\, v_e^3}{n_e\, \varepsilon^2\, \alpha\, \alpha_D}\,,
~}
\label{eq:t_dark_from_SM}
\end{align}
valid in the Coulomb-dominated regime (appropriate for Planck relics). Thus $t_{\rm neut}^{(D\leftarrow {\rm SM})}\propto \varepsilon^{-2}$: even modest ionized SM environments neutralize $Q_D$ rapidly unless $\varepsilon$ is extremely small. Requiring $t_{\rm neut}^{(D\leftarrow {\rm SM})}\gtrsim H_0^{-1}$ in the ionized IGM gives the rough bound
\begin{align}
\varepsilon \;\lesssim\; \Bigg[\frac{\pi}{24}\,
\frac{m_e^2\, v_e^3}{\alpha\, \alpha_D\, n_e\, H_0^{-1}}\Bigg]^{1/2},
\end{align}
which is typically far below existing laboratory and astrophysical limits for $\alpha_D\!\sim\!\alpha$; hence, achieving Hubble-scale protection via mixing alone generally requires \emph{ultra}-small $\varepsilon$.

\subsection{Massive dark photon (Yukawa field)}
\label{subsec:yukawa}

If $U(1)_D$ is broken and the dark photon has mass $m_{A'}$, the exterior field is Yukawa-suppressed \cite{Okun:1982xi,Galison:1983pa,Jaeckel:2010ni,Fabbrichesi:2020wbt}, $E_D(r)\sim (Q_D/4\pi r^2)\,(1+m_{A'}r)\,e^{-m_{A'} r}$. Two consequences follow:

\begin{enumerate}
\item \textbf{Ambient capture saturates at the force range.}
The focusing radius cannot exceed the force range $\lambda_D'\equiv m_{A'}^{-1}$; hence the geometric cross section saturates at $\sigma\lesssim \pi \lambda_D'^2$, implying the conservative lower bound $t_{\rm neut}\gtrsim v/(n\,\pi \lambda_D'^2)$ when $r_{\rm acc}>\lambda_D'$. For Planck relics and any $m_{A'}\!\ll\!M_{\rm Pl}$, the near zone is Coulombic ($r_h\ll \lambda_D'$), so the unscreened formula \eqref{eq:t_dark_coulomb} applies unless $m_{A'}$ is \emph{very} large.

\item \textbf{Dark Schwinger is controlled by the near zone.}
Since $r_h\ll \lambda_D'$ for any realistic $m_{A'}$, the horizon field is unaffected by the Yukawa tail and the threshold $e_D E_D(r_h)\sim m_D^2$ is unchanged. Thus eq.~\eqref{eq:Qstop_dark} remains valid.
\end{enumerate}

\paragraph{Direct loss of charge for a massive $U(1)_D$.}
Note that if the dark photon is massive ($m_{A'}>0$), there is no Gauss-law-protected charge and the exterior Proca field relaxes on a timescale $t_{\rm loss}\sim m_{A'}^{-1}$ even in vacuum. This is the standard result that massive spin-1 “hair” decays: when a source falls through the horizon the exterior field leaks away in Schwarzschild time $\sim \mu^{-1}$ and persists indefinitely only in the $\mu\!\to\!0$ limit~\cite[Sec.~2; discussion around the massive vector case]{Coleman:1991QuantumHair}. Quantitatively, retaining observable dark charge over a Hubble time requires $m_{A'}\lesssim H_0\!\simeq\!1.5\times10^{-33}\,{\rm eV}$; any \emph{appreciable} breaking ($m_{A'}\!\gg\!H_0$) erases the hair on microscopic timescales, independently of ambient capture or Schwinger discharge. Thus a Higgsed $U(1)_D$ cannot sustain cosmological dark charge unless it leaves an exact remnant (e.g.\ a discrete gauge subgroup), in which case only exponentially suppressed, non-perturbative ``quantum hair'' survives without a long-range field~\cite{Coleman:1991QuantumHair}.

\subsection{Regime map and benchmarks}
\label{subsec:dark_regimes}

Combining \eqref{eq:Qdot_dark_exact}, \eqref{eq:t_dark_trigamma}-\eqref{eq:t_dark_coulomb}, and \eqref{eq:t_dark_from_SM}, the fate of a dark-charged Planck relic is controlled by:

\begin{itemize}
\item \textbf{Dark Schwinger (vacuum):} operative only if $e_D E_D(r_h)\gtrsim m_D^2$; if not, $Q_D$ is stable to vacuum discharge and remains at its initial value up to ambient effects; when operative, discharge halts at $Q_{D,\,\rm stop}\propto (m_D^2/\alpha_D)\,r_h^2$.

\item \textbf{Ambient dark capture:} timescale $t_{\rm neut}^{(D)}\propto (m_D^2/\alpha_D^2)\,v_D^3/n_D$; a cold, symmetric dark plasma neutralizes quickly. If the symmetric component is depleted ($n_D\!\to\!0$), this channel is absent.

\item \textbf{Kinetic-mixing leakage to the SM:} timescale $t_{\rm neut}^{(D\leftarrow {\rm SM})}\propto (\varepsilon^2 \alpha \alpha_D)^{-1}\, m_e^2 v_e^3/n_e$; unless $\varepsilon$ is extremely small, ionized SM environments (reionization, IGM, ISM) neutralize $Q_D$ rapidly.
\end{itemize}

%\noindent
%\textbf{Benchmarks.} For illustration, set $\alpha_D=\alpha$, $m_D=10^{10}\,\mathrm{GeV}$, a Planck relic ($r_h\simeq \ell_{\rm Pl}$). Then $Q_{D,\,\rm stop}/e \sim (m_D/m_e)^2 \sim 10^{34}$, i.e.\ vacuum discharge never turns on unless the relic is born with enormous $Q_D$. If the dark plasma is negligible ($n_D\!\to\!0$), neutralization proceeds entirely through kinetic-mixing leakage with timescale $t_{\rm neut}^{(D\leftarrow {\rm SM})}\propto \varepsilon^{-2}$; requiring $t\gtrsim H_0^{-1}$ in the ionized IGM typically demands $\varepsilon\!\ll\!10^{-12}$-$10^{-14}$ (order-of-magnitude, depending on $n_e$ and $v_e$), well beyond most motivated ranges. Conversely, a modest symmetric dark plasma with $n_D\!\gtrsim\!10^{-10}\,n_b$ and $v_D\!\lesssim\!10^{-3}$ already makes $t_{\rm neut}^{(D)}$ sub-cosmological even for large $m_D$.

\section{No-Schwinger-pair limit: conditions for enduring SM electric charge}
\label{sec:noSch}

Assume now that Planck-scale new physics \emph{completely} shuts off QED vacuum pair creation near the horizon (Sec.~\ref{sec:schwinger}), so that neutralization can proceed only through \emph{ambient capture} of opposite charges in plasmas or neutral media (Sec.~\ref{sec:capture}). Let me now derive conditions on a relic’s mass \(M\) and SM charge \(Q=Ze\) such that \emph{no discharge occurs} over cosmological history.

\subsection{Global ``no-capture'' condition}
\label{subsec:nocap}

The capture rate for opposite charges in a medium with number density \(n_{\rm ch}(t)\) and characteristic speed \(v(t)\) is (eq.~\eqref{eq:sigma_general})
\begin{align}
\dot N_k(t) \;=\; \frac{4\pi\,n_{\rm ch}(t)}{v(t)^3}\,\Big(GM+\kappa\,k\Big)^2,
\qquad \kappa \equiv \frac{e^2}{4\pi\epsilon_0\,m_c},
\end{align} when the relic’s instantaneous charge is \(k e\) and the captured species has mass \(m_c\)
(\(m_c\!=\!m_e\) for electrons if the relic is positive, \(m_c\!=\!m_p\) for protons if negative). If \emph{no capture at all} is to occur, it must in particular be true that the expected number of captures while the relic sits at its \emph{initial} charge \(k=|Z|\) is \(<1\). This yields a necessary (and very nearly sufficient) condition
\begin{align}
\boxed{\;
\mathcal{I}_{\rm ch}\;\equiv\; 4\pi\,\!\int_{t_{\rm form}}^{t_0} \! dt\, \frac{n_{\rm ch}(t)}{v(t)^3}\;
\Big(GM+\kappa\,|Z|\Big)^2 \;<\; 1\,.
\;}
\label{eq:global_condition}
\end{align}
It is convenient to define the \emph{cosmic exposure}
\begin{align}
\mathcal{E}_{\rm ch} \;\equiv\; \int_{t_{\rm form}}^{t_0}\! dt\;\frac{n_{\rm ch}(t)}{v(t)^3},
\end{align}
so that the no-capture condition is simply
\begin{align}
\boxed{\;
GM+\kappa\,|Z| \;<\; \frac{1}{\sqrt{4\pi\,\mathcal{E}_{\rm ch}}}\,.
\;}
\label{eq:master_ineq}
\end{align}
Accounting for a possible Planck-scale suppression of horizon absorption simply rescales the exposure as $\mathcal{E}\!\to\!\beta_{\rm abs}\,\mathcal{E}$.

Two useful limits follow immediately:
\begin{align}
\text{(Coulomb-dominated; }GM\ll \kappa |Z|) &:~~
\boxed{~|Z| \;<\; \frac{1}{\kappa}\,\frac{1}{\sqrt{4\pi\,\mathcal{E}_{\rm ch}}}~,}
\label{eq:Zmax_general}\\[4pt]
\text{(Gravity-dominated; }GM\gg \kappa |Z|) &:~~
\boxed{~M \;<\; \frac{1}{G}\,\frac{1}{\sqrt{4\pi\,\mathcal{E}_{\rm ch}}}~.}
\label{eq:Mmax_general}
\end{align}
These inequalities must hold for \emph{each} relevant charge carrier (electrons for \(Z>0\), protons for \(Z<0\)); the stronger of the two constraints applies for any charged relics to exist in the late universe.

\paragraph{Screening saturation.} If at some epoch the focusing radius exceeds the Debye length \(\lambda_D\), the cross section saturates at \(\sigma\!\sim\!\pi\lambda_D^2\), replacing \( (GM+\kappa k)^2/v^4 \mapsto \min\!\big[(GM+\kappa k)^2/v^4,\,\lambda_D^2\big]\) under the time integral. In cosmological and galactic plasmas one typically has \(r_{\rm acc}\ll \lambda_D\), so eqs.~\eqref{eq:global_condition}-\eqref{eq:Mmax_general} are conservative.

\subsection{Conservative numerical bounds (late Universe only)}
\label{subsec:conservative}

Even without integrating over the dense early Universe, the \emph{late} ionized IGM (post-reionization) already forces extremely strong bounds. Taking representative values today
\[
n_e \simeq 0.25~\mathrm{m^{-3}},\qquad
v_e \simeq \sqrt{\frac{8k_B T}{\pi m_e}} \sim 6.2\times 10^5~\mathrm{m\,s^{-1}}~~(T\sim 10^4~\mathrm{K}),
\]
over a Hubble time \(t_0\!\simeq\!4.35\times 10^{17}\) s gives
\(\mathcal{E}_e \simeq \int dt\, n_e/v_e^3 \approx 0.45\).
With \(\kappa_e\!\equiv\!e^2/(4\pi\epsilon_0 m_e)\!\simeq\!253~\mathrm{m^3\,s^{-2}}\) and \(\kappa_p\!\equiv\!e^2/(4\pi\epsilon_0 m_p)\!\simeq\!0.138~\mathrm{m^3\,s^{-2}}\), eq.~\eqref{eq:master_ineq} yields:

\begin{itemize}
\item \textbf{Positive relics (electron capture; Coulomb-dominated).}
Even if \(M\!\to\!0\), the no-capture bound \eqref{eq:Zmax_general} gives
\[
|Z| \;<\; \frac{1}{\kappa_e\sqrt{4\pi\mathcal{E}_e}}
\;\approx\; 1.6\times 10^{-3}\,,
\]
impossible for quantized charge. Thus a positively charged relic will capture at least one electron from the present-day IGM alone \emph{within a Hubble time}.

\item \textbf{Negative relics (proton capture; typically gravity- or mixed-dominated).}
Using \(\kappa_p\) gives a weaker constraint today: \(|Z| \lesssim 3\). However, including only the reionization era (\(n_e\!\sim\!10^2~\mathrm{m^{-3}}\), \(T\!\sim\!10^4\) K, duration \(\sim\!0.5\) Gyr) adds \(\Delta\mathcal{E}_e \sim 7\). Then \(4\pi\mathcal{E}_e \gtrsim 90\) and \(|Z| \lesssim 0.8\), again excluding any nonzero integer \(Z\).
\end{itemize}

\noindent
These are \emph{conservative} because they ignore neutralization before and during reionization, where densities were higher (even though \(v\) was also larger).

\smallskip\noindent
Numerically, using the post-reionization IGM ($\mathcal{E}_e\simeq 0.45$ for electrons), one finds that avoiding capture of even a single electron requires $\beta_{\rm abs}$ suppressed by {\it six-seven orders of magnitude} relative to geometric focusing. I discuss this in detail in Sec.~\ref{subsec:conclude_nosch} below. Including the reionization exposure tightens this bound further. For $Z<0$ (proton capture), the classical condition already excludes $|Z|\!\ge\!1$ once reionization is accounted for; only an $\mathcal{O}(1)$ greybody factors suppression would reopen that window. Importantly, detailed balance links greybody absorption and Hawking emission, so any mechanism that drives $\beta_{\rm abs}\!\to\!0$ (turning off evaporation) generically suppresses absorption by the same factor. Thus, unless charged greybody factors are extraordinarily small ($\beta_{\rm abs}\!\ll\!10^{-6}$), the focusing arguments remain robust.

\subsection{Mass bounds in the gravity-dominated regime}
\label{subsec:massbounds}

When \(GM\) dominates over \(\kappa |Z|\) (very heavy relics and/or negative relics in proton media), the no-capture bound \eqref{eq:Mmax_general} with \(\mathcal{E}_e \gtrsim 0.45\) gives
\[
M \;<\; \frac{1}{G\sqrt{4\pi\mathcal{E}_e}} \;\approx\; 6\times 10^9~\mathrm{kg}.
\]
Adding only reionization (\(\Delta\mathcal{E}_e\!\sim\!7\)) tightens this to \(M \lesssim 1.6\times 10^9\) kg. Any heavier relic will \emph{certainly} capture at least one opposite charge over late-time cosmic history, even with vacuum Schwinger entirely shut off.

\subsection{Conclusion for SM electromagnetism}
\label{subsec:conclude_nosch}

Equations~\eqref{eq:Zmax_general}-\eqref{eq:Mmax_general} provide the precise no-discharge conditions when
vacuum pair production is absent. Using only post-reionization exposures already implies:
\begin{itemize}
\item No positively charged relic with integer \(|Z|\) can avoid capturing an electron.
\item No negatively charged relic can keep \(|Z|\ge 1\) once the reionization exposure is included.
\item Increasing \(M\) worsens the problem (gravity strengthens capture); decreasing \(M\) does not help
because the Coulomb term dominates and the bound on \(|Z|\) remains \(<1\).
\end{itemize}

Let us define a parameter $\epsilon$ to quantify the fractional  distance of a Reissner--Nordström black hole from exact extremality. In SI units $\epsilon$ is defined as 
\begin{equation}
\epsilon \;\equiv\; \frac{M c^2 - |Q| \sqrt{G/(4\pi \epsilon_0)}}{M c^2}.
\end{equation}
Thus $\epsilon=0$ corresponds to an exactly extremal black hole,  while larger $\epsilon$ denotes increasing departure from extremality. In Planck units ($G=\hbar=c=4\pi\epsilon_0=1$), the expression simplifies to $\epsilon = 1 - |Q|/M$. 
\begin{figure}[t]
    \centering
    \includegraphics[width=0.5\textwidth]{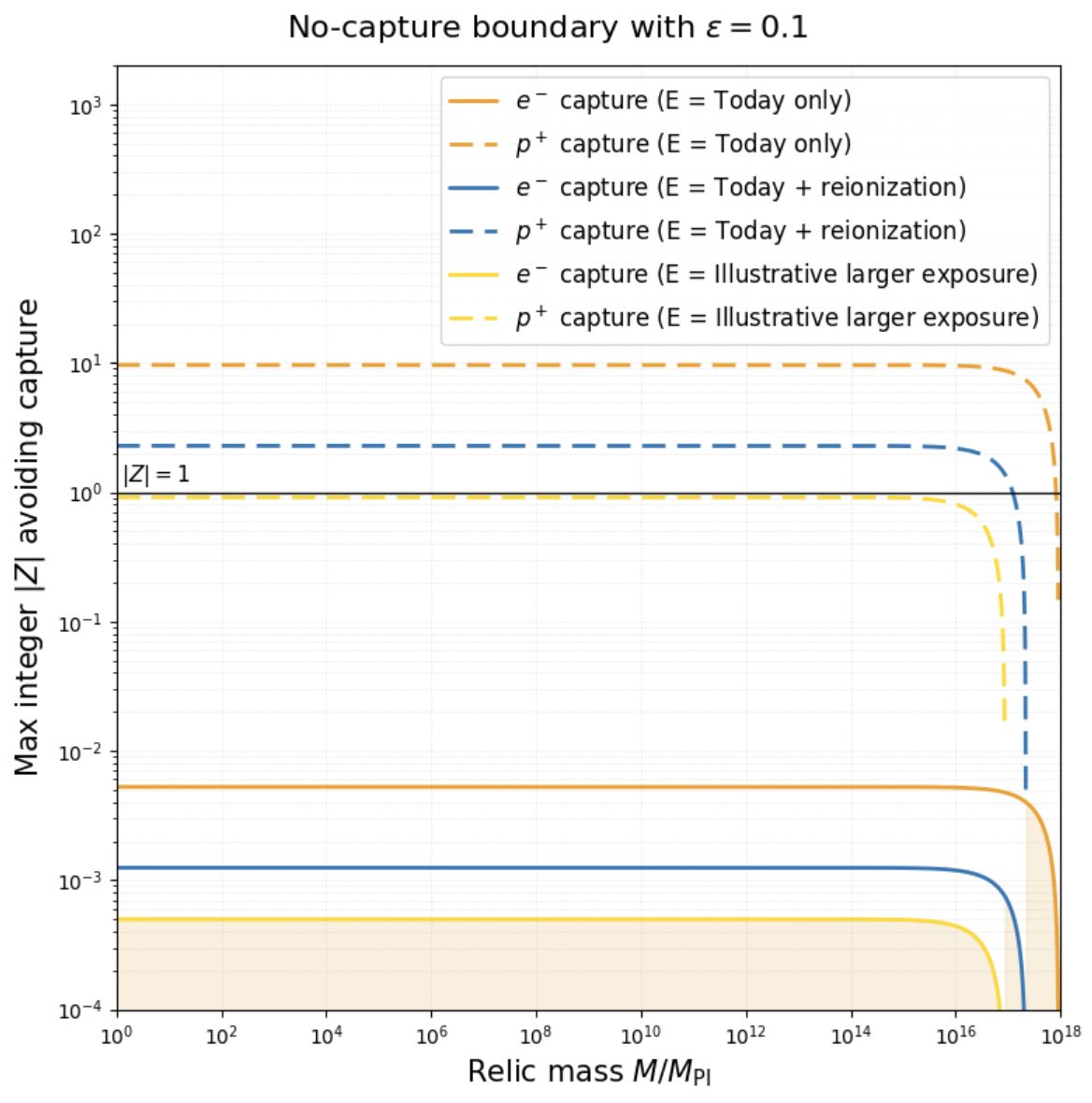}
    \caption{\textbf{No-capture boundary with extremality suppression $\epsilon=0.1$.}
    Shown is the maximum integer relic charge $|Z|$ that avoids Coulomb-focused capture of ambient electrons (solid) or protons (dashed) as a function of relic mass. 
    Three effective exposures are considered: today only ($\mathcal{E}=0.45$), today plus reionization ($\mathcal{E}=8$), and an illustrative larger exposure ($\mathcal{E}=50$), each uniformly suppressed by $\beta_{\rm abs}=\epsilon=0.1$. 
    The black horizontal line marks $|Z|=1$. 
    Shaded regions below the lowest electron curve are excluded, as any relic in that range would have been neutralized across cosmic history.}
    \label{fig:no_capture_eps01}
\end{figure}
 Figure~\ref{fig:no_capture_eps01} illustrates the impact of imposing an extremality suppression factor  $\epsilon = 0.1$ on the no-capture condition. 
I assume that Schwinger pair production is switched off, so vacuum discharge does not operate and  the constraints arise purely from ambient capture. Three benchmark exposures are considered: today only  ($\mathcal{E}=0.45$), today plus reionization ($\mathcal{E}=8$), and an illustrative larger exposure  ($\mathcal{E}=50$), where $\mathcal{E}$ encodes the integrated density and velocity distribution of  charges encountered over cosmic time. Reducing the effective exposures by the suppression factor $\epsilon$ raises the no-capture boundaries in $|Z|$, apparently allowing larger charges to survive at  high mass. Nevertheless, even with $\epsilon=0.1$ and the most conservative exposure, electron capture  enforces $|Z|\lesssim 10^{-3}$ across essentially the entire mass range, excluding all integer charges unless absorption is suppressed by many further orders of magnitude. Proton capture is somewhat less restrictive, but still rules out $|Z|\geq 1$ once realistic exposures are included. This example shows that moderate extremality suppression, in the absence of Schwinger discharge, is insufficient: avoiding neutralization would require relics to remain extremely close to extremality and to invoke additional new physics to suppress absorption well beyond the $\epsilon=0.1$ case.

\begin{figure}[t]
    \centering
    \includegraphics[width=\textwidth]{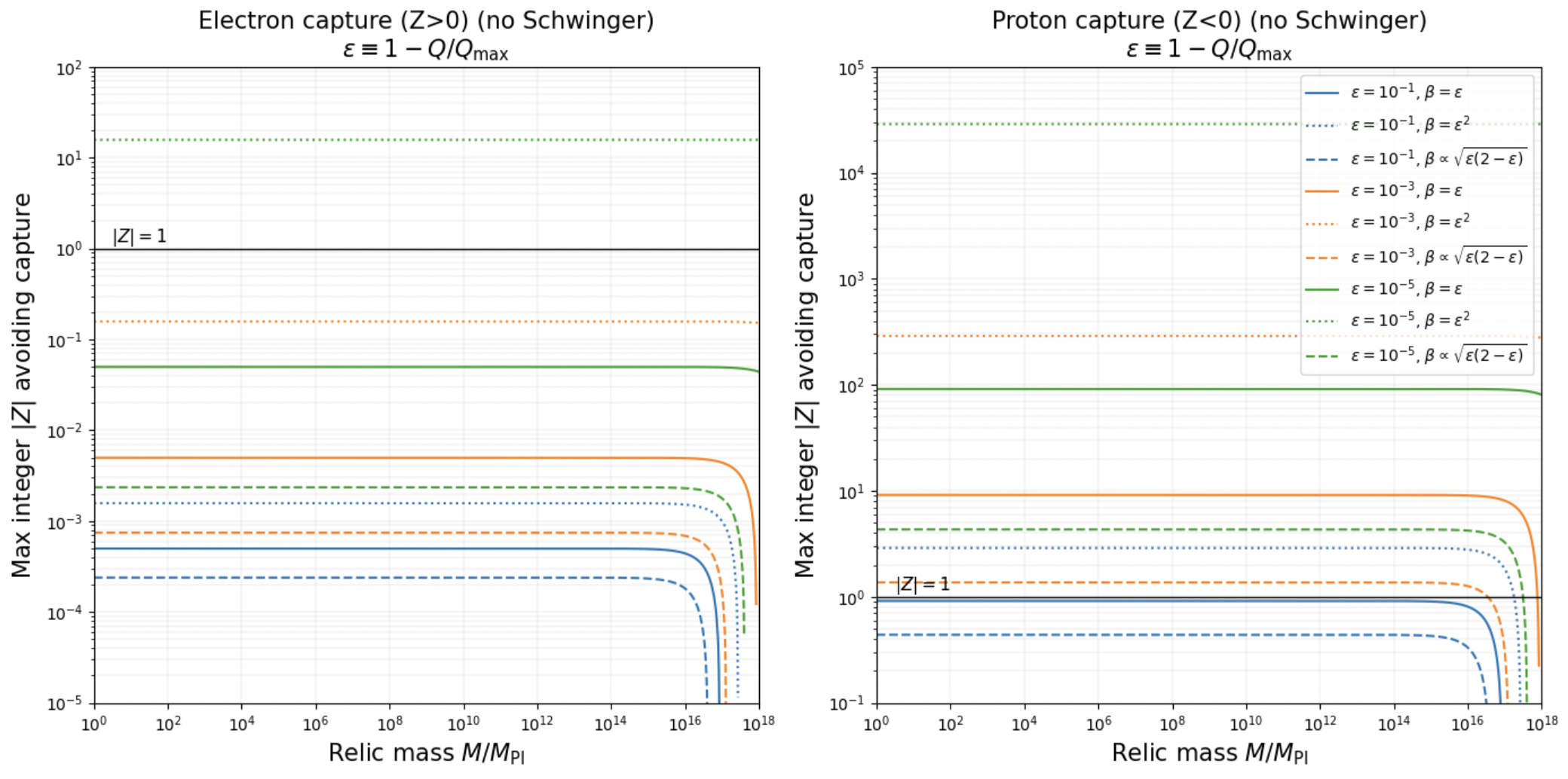}
    \caption{\textbf{No-capture boundaries with extremality suppression, split by charge sign.} 
    Left: electron capture ($Z>0$). Right: proton capture ($Z<0$). 
    In both panels Schwinger discharge is switched off, and I assume $\mathcal{E}=50$. 
    The parameter $\epsilon \equiv 1 - Q/Q_{\max}$ quantifies proximity to extremality, 
    and different mappings $\beta_{\rm abs}(\epsilon)$ are shown: $\beta=\epsilon$ (solid), 
    $\beta=\epsilon^2$ (dotted), and $\beta\propto\sqrt{\epsilon(2-\epsilon)}$ (dashed). 
    Each color corresponds to a different value of $\epsilon$: $10^{-1}$ (blue), $10^{-3}$ (orange), and $10^{-5}$ (green). 
    The black line marks $|Z|=1$.}
    \label{fig:no_capture_epsilon01}
\end{figure}

\smallskip\noindent
Figure~\ref{fig:no_capture_epsilon01}  shows the no-capture boundaries for positively charged relics  (left panel, electron capture) and negatively charged relics (right panel, proton capture), under the  assumption that Schwinger discharge is switched off and that the effective cosmic exposure is fixed at  $\mathcal{E}=50$. This value of $\mathcal{E}$ corresponds to an upper bound on the integrated  encounter with ambient charges across cosmic history, so the constraints plotted here are already  maximally conservative. 

\smallskip\noindent
The panels explore how different prescriptions for the absorption suppression factor,  $\beta_{\rm abs}(\epsilon)$, modify the neutralization boundaries. The functional form of $\beta_{\rm abs}(\epsilon)$ is unknown, and model dependent. Three functional forms are therefore compared: a linear mapping $\beta=\epsilon$ (solid), a quadratic mapping $\beta=\epsilon^2$ (dotted), and a square-root form $\beta \propto \sqrt{\epsilon(2-\epsilon)}$ (dashed). For each, results are shown at three benchmark values of the extremality parameter, $\epsilon = 10^{-1}, 10^{-3}, 10^{-5}$. 

\noindent
\smallskip\noindent
Electron capture (left) remains the most constraining case: for all $\beta_{\rm abs}$ prescriptions,  the allowed charge is driven far below $|Z|=1$, excluding integer charges across the full mass range. Proton capture (right) is less restrictive, with some survival of $|Z|\gtrsim 1$ at large masses depending on the choice of $\beta_{\rm abs}$. Still, the differences among the mappings are modest, and even the most optimistic suppression scenarios leave only a narrow window where integer charges could persist. 

\smallskip\noindent
Overall, the figure demonstrates that once realistic exposures are saturated at $\mathcal{E}=50$,  the precise choice of $\beta_{\rm abs}(\epsilon)$ controls the detailed shape of the boundaries, but does not qualitatively alter the outcome: integer-charged relics are generically excluded, while only fractional charges $|Z|\ll 1$ remain compatible with long-term survival.

\smallskip\noindent
Including gravitational focusing sharpens the no-capture boundaries: while at low and moderate 
masses Coulomb focusing by ambient electrons or protons controls the discharge limits, at 
sufficiently large masses the gravitational term dominates and enforces neutralization even when 
the electric charge is tiny. In practice this means that the Coulomb discharge curves terminate 
in a vertical asymptote beyond which gravity alone ensures capture; all relic masses to the right 
of that line are automatically excluded. For the maximal exposure $\mathcal{E}=50$, the 
gravitational thresholds occur at the following values:  
\begin{itemize}
  \item For $\epsilon=10^{-1}$: 
    \begin{itemize}
      \item $\beta=\epsilon$: $M_{\rm crit}^{\rm grav}\simeq 6\times 10^{9}\,\mathrm{kg} 
        \;\;(\sim 3\times 10^{17}\,M_{\rm Pl})$,
      \item $\beta=\epsilon^2$: $M_{\rm crit}^{\rm grav}\simeq 6\times 10^{10}\,\mathrm{kg} 
        \;\;(\sim 3\times 10^{18}\,M_{\rm Pl})$,
      \item $\beta=\sqrt{\epsilon(2-\epsilon)}$: $M_{\rm crit}^{\rm grav}\simeq 1.4\times 10^{9}\,\mathrm{kg} 
        \;\;(\sim 6\times 10^{16}\,M_{\rm Pl})$.
    \end{itemize}
  \item For $\epsilon=10^{-3}$: 
    \begin{itemize}
      \item $\beta=\epsilon$: $M_{\rm crit}^{\rm grav}\simeq 6\times 10^{11}\,\mathrm{kg} 
        \;\;(\sim 3\times 10^{19}\,M_{\rm Pl})$,
      \item $\beta=\epsilon^2$: $M_{\rm crit}^{\rm grav}\simeq 6\times 10^{14}\,\mathrm{kg} 
        \;\;(\sim 3\times 10^{22}\,M_{\rm Pl})$,
      \item $\beta=\sqrt{\epsilon(2-\epsilon)}$: $M_{\rm crit}^{\rm grav}\simeq 1.3\times 10^{10}\,\mathrm{kg} 
        \;\;(\sim 6\times 10^{17}\,M_{\rm Pl})$.
    \end{itemize}
  \item For $\epsilon=10^{-5}$: 
    \begin{itemize}
      \item $\beta=\epsilon$: $M_{\rm crit}^{\rm grav}\simeq 6\times 10^{13}\,\mathrm{kg} 
        \;\;(\sim 3\times 10^{21}\,M_{\rm Pl})$,
      \item $\beta=\epsilon^2$: $M_{\rm crit}^{\rm grav}\simeq 6\times 10^{18}\,\mathrm{kg} 
        \;\;(\sim 3\times 10^{26}\,M_{\rm Pl})$,
      \item $\beta=\sqrt{\epsilon(2-\epsilon)}$: $M_{\rm crit}^{\rm grav}\simeq 1.3\times 10^{11}\,\mathrm{kg} 
        \;\;(\sim 6\times 10^{18}\,M_{\rm Pl})$.
    \end{itemize}
\end{itemize}
Thus, once gravitational capture is accounted for, there is no viable charged-relic window at 
masses larger than these thresholds, regardless of the Coulomb suppression mechanism.

\smallskip\noindent
The exclusion of integer charges naturally raises the question of whether relics could instead survive with \emph{fractional} charges, $|Z|\ll 1$, in the form of effectively millicharged dark matter. Such scenarios have been widely studied in the literature. A tiny electric charge can arise through kinetic mixing between the photon and a hidden $U(1)$ gauge boson, through Stueckelberg mechanisms, or from embedding electromagnetism in a larger gauge structure where certain dark-sector states carry non-integer charges under the residual $U(1)_{\rm EM}$. In these constructions, the lightest dark-sector particles can acquire charges as small as $q_\chi \sim 10^{-6}e$ or below, while still interacting gravitationally as viable DM candidates. 

\smallskip\noindent
Millicharged dark matter has been invoked in a wide range of contexts, from explanations of the EDGES 21-cm anomaly to models of self-interacting DM with suppressed electromagnetic couplings. Comprehensive reviews and updated constraints can be found in Refs.~\cite{Davidson:2000hf,Vogel:2013raa,Jaeckel:2010ni}, which cover theoretical origins, cosmological and astrophysical limits, and experimental search strategies (including fixed-target experiments, collider probes, and precision tests of Coulomb’s law). For our purposes, the salient  point is that millicharges evade the capture bounds discussed above: relics with $|Z|\ll 1$ fall  comfortably below the no-capture thresholds, and are not driven to neutralization even when  $\mathcal{E}$ saturates its maximal value.

\smallskip\noindent
In conclusion, generically \emph{even if Schwinger discharge is completely shut off by Planck-scale new physics}, Standard Model-charged relics cannot remain charged over cosmological times. The only exception is the limiting case of an exactly extremal Reissner-Nordström black hole, corresponding to $\epsilon \to 0$, where the absorption coefficient  vanishes. In this regime, macroscopic charged relics do not neutralize through ambient capture, since infalling charges are no longer efficiently absorbed. Outside this finely tuned extremal limit, however, relics are neutralized whenever light opposite charges are present in the plasma. The only other viable pathways to evade discharge are those identified in Sec.~\ref{sec:newphysics}: relocating the charge to a hidden sector with no light opposite carriers (e.g.~a $U(1)_D$ with a heavy lightest charged state and/or extremely small kinetic mixing), or endowing the relic with a discrete or topological charge rather than an ordinary Standard Model electric charge.

\section{Discussion and conclusions}
\label{sec:conclusions}

The exterior-field physics analyzed here shows that, under Standard Model dynamics,  Planck relics carrying ordinary electric charge cannot retain it.  

\begin{itemize}
\item \textbf{Ambient capture} neutralizes relics rapidly in all relevant environments.  The exact sum over discharge steps, eqs.~\eqref{eq:sigma_general} and \eqref{eq:t_trigamma}, reduces to the Coulomb-dominated limit \eqref{eq:t_coulomb} for Planck relics and is independent of $M$. Across cosmic epochs---MeV era, recombination, reionization, and the present-day IGM/ISM---the resulting times are orders of magnitude shorter than Hubble times 
(Table~\ref{tab:epochs}).  
\item \textbf{Vacuum discharge} alone enforces neutrality in the absence of plasma.  Integrating the local Schwinger kernel yields the explicit discharge law (Sec.~\ref{sec:schwinger}); discharge proceeds until $E(r_h)\simeq E_{\rm crit}$, leaving a residual charge $Q_{\rm stop}^{(e)}\propto r_h^2\ll e$ for Planck radii.  
\item \textbf{Mass systematics} reinforce these conclusions. For $M\gg M_\star$ gravity accelerates capture ($t_{\rm neut}\propto M^{-2}$ at fixed $Z$), while Schwinger discharge remains kinematically available up to $M\sim 10^{36}\,$kg, i.e. in the supermassive BH mass range. Sub-Planckian compact objects (no horizon) neutralize even more efficiently.  
\item \textbf{Quantum gravity} does not offer a loophole: exterior fields obey Gauss’s law,  semiclassical QED is reliable on the production shell, and plausible UV deformations (running $\alpha$, Euler-Heisenberg/Born-Infeld-type corrections) are too small to alter the picture. Swampland-type arguments (e.g.\ the Weak Gravity Conjecture) in fact favor additional discharge channels.  
\item \textbf{Even without Schwinger pair production}, cosmological exposure alone excludes stable SM charge. The global no-capture condition  $(GM+\kappa|Z|)\sqrt{4\pi\mathcal{E}}<1$ forbids any integer $|Z|$ for positive relics based only on today’s IGM, and rules out negative relics once reionization is included.  
\end{itemize}

\noindent
Taken together, these results imply that SM-charged relics neutralize efficiently in all cosmological settings. Three possible loopholes remain:  
\begin{enumerate}
\item A highly fine-tuned Standard Model corner in which (i) the relic sits arbitrarily close to Reissner-Nordström extremality---a possibility we discuss in App.~\ref{sec:app}---so that all greybody factors for charged absorption are suppressed, and (ii) SM Schwinger pair creation is completely absent due to Planck-scale new physics.  Only in this contrived double limit can ordinary discharge channels be evaded.  
\item Relocating the charge to a hidden $U(1)_D$ sector with no light opposite charge carriers, for example if the lightest charged state is very heavy and/or the kinetic mixing with $U(1)_{\rm EM}$ is extremely suppressed.  
\item Endowing the relic with a discrete or topological gauge charge rather than an ordinary SM electric charge.  
\end{enumerate}

\noindent In summary, if black holes, including Planck relics, and up to supermassive black holes, compose dark matter or a fraction thereof, they must be  \emph{electrically neutral} with respect to the SM. This work generalizes the Gibbons–MacGibbon discharge analyses by incorporating environmental capture, cosmological exposure, and dark-sector extensions, demonstrating that electrically charged relics cannot survive within Standard-Model electromagnetism. Only the three loopholes above---one fine-tuned within the SM, and two involving extensions thereof---offer viable ways to retain charge.   Thus, except in such 
contrived scenarios, \emph{charged} primordial black holes, including their evaporation relics, neutralize.

\acknowledgments
This work is partly supported by the U.S.\ Department of Energy grant number de-sc0010107; I am grateful to Benjamin V. Lehmann and Thomas Schwemberger for important feedback and suggestions on this manuscript.

\appendix

\section{Extremality and Evaporation in the Reissner--Nordström Case}\label{sec:app}

The late-time behavior of evaporating Reissner-Nordstr\"om (RN) black holes is controlled by the competition between charge-dependent modifications to the Hawking spectrum, stochastic fluctuations in emission, and the suppression of the temperature as extremality is approached. Understanding whether charged relics can persist therefore requires tracking how the evaporation trajectory evolves with both mass and time. Near-extremal configurations are especially significant, since the approach toward extremality alters the thermal history and leads to evaporation rates that slow dramatically, without ever halting completely in finite time.

\subsection{Temperature near extremality}

\begin{figure}[t]
  \centering
  \begin{minipage}[t]{0.48\textwidth}
    \centering
    \includegraphics[width=\textwidth]{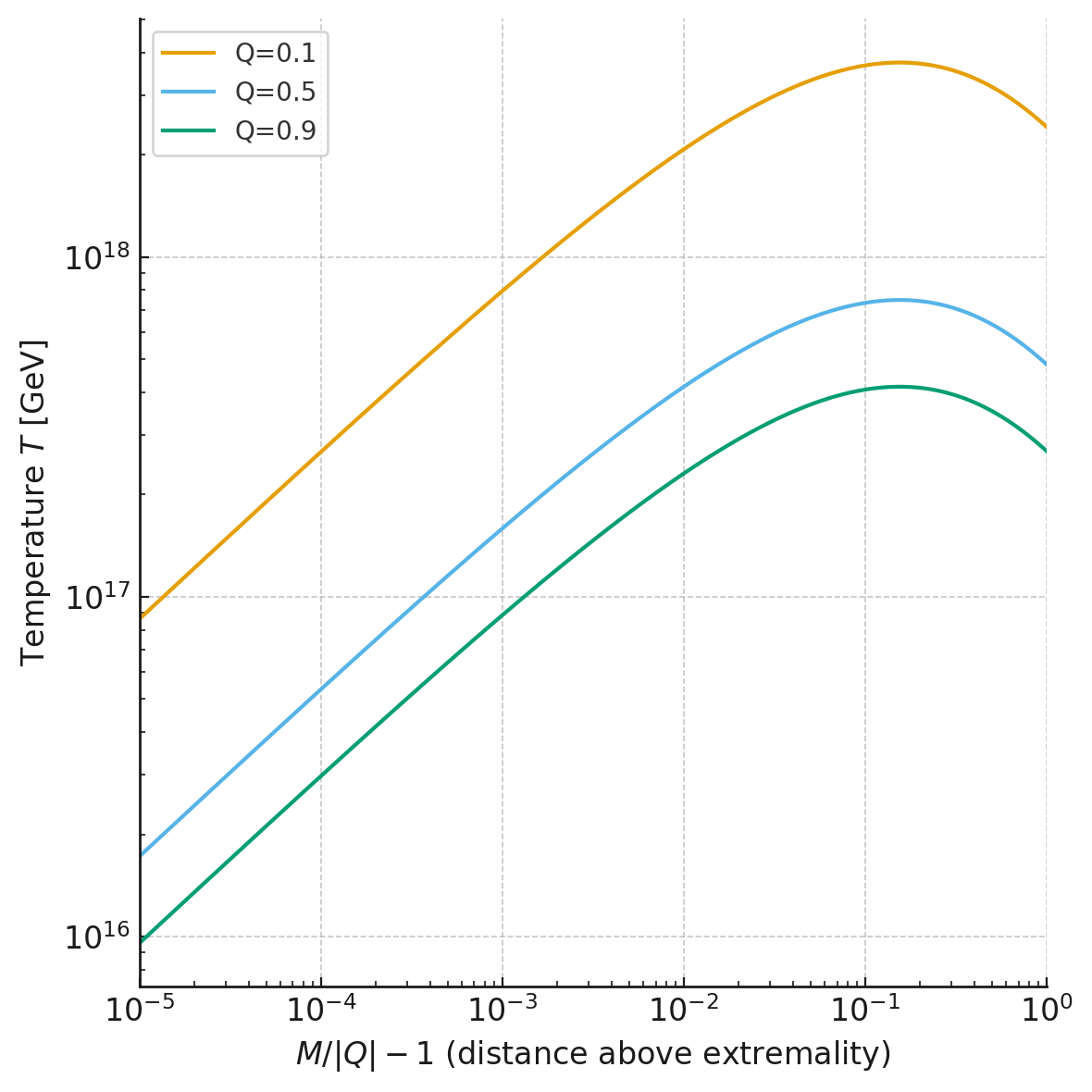}
  \end{minipage}\hfill
  \begin{minipage}[t]{0.48\textwidth}
    \centering
    \includegraphics[width=\textwidth]{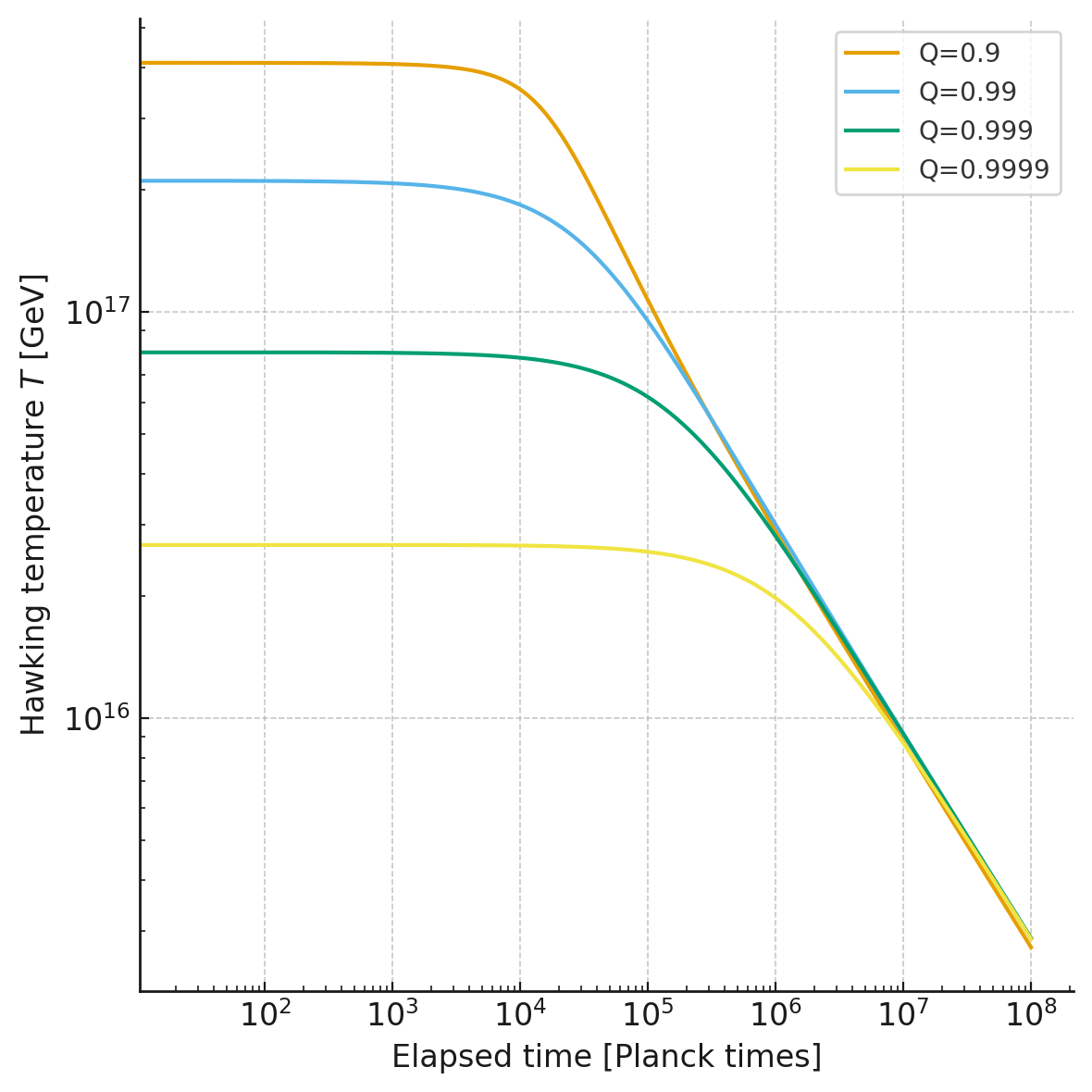}
  \end{minipage}
  \caption{
  {\bf Left:} Hawking temperature $T$ of Reissner--Nordström black holes as a function of the 
  distance from extremality, $\delta \equiv M/|Q|-1$, for charges $Q=0.1,\,0.5,\,0.9$. 
  The curves illustrate the generic behavior: $T\to 0$ at extremality, a $\sqrt{\delta}$ rise, 
  a maximum at intermediate $\delta$, and Schwarzschild-like cooling at large mass. 
  {\bf Right:} Time evolution of the Hawking temperature $T$ for near-extremal RN black holes 
  with charges $Q=0.9,\,0.99,\,0.999,\,0.9999$. After an early plateau, all trajectories 
  converge to the universal $T \propto t^{-1/2}$ decay, reflecting the asymptotic approach 
  to extremality. Together, the two panels summarize the mass-domain and time-domain 
  perspectives on how evaporation drives RN black holes toward extremality.
  }
  \label{fig:T_vs_delta_time}
\end{figure}

Figure~\ref{fig:T_vs_delta_time}, left, shows the Hawking temperature of RN black holes as a function of the fractional distance from extremality, 
\begin{equation}
  \delta \equiv \frac{M}{|Q|} - 1,
\end{equation}
for representative charges $Q = 0.1, 0.5, 0.9$ in Planck units. All three cases display the characteristic non-monotonic behavior: the temperature vanishes at exact extremality ($\delta \to 0$), rises as $T \propto \sqrt{\delta}$ close to extremality, reaches a maximum at intermediate $\delta$, and eventually falls off as $T \propto 1/M$ in the Schwarzschild regime ($M \gg Q$). The location of the turnover depends on $Q$: smaller charges yield higher peak temperatures and shift the maximum to larger $\delta$, while larger charges suppress the temperature scale and push the maximum closer to extremality. This pattern highlights that charged black holes generically spend a finite portion of their evolution at elevated temperatures before cooling again as the neutral limit is approached.

\subsection{Time evolution toward extremality}

Figure~\ref{fig:T_vs_delta_time}, right, complements this picture by showing the temperature as a function of elapsed time (again measured in Planck units) for black holes with $Q = 0.9, 0.99, 0.999, 0.9999$. The initial condition in all cases is $M = M_{\rm Pl}$. The evolution exhibits a short-lived plateau phase, during which the temperature is nearly constant, followed by the universal scaling regime
\begin{equation}
  T(t) \;\propto\; t^{-1/2},
\end{equation}
which arises from the near-extremal dynamics $d\delta/dt \propto -\delta^2$ and $T \propto \sqrt{\delta}$. This behavior demonstrates that, once sufficiently close to extremality, RN black holes cool asymptotically, with evaporation rates that slow dramatically and prevent the system from ever reaching the exactly extremal state in finite time. The precise value of $Q$ determines the height of the initial plateau, but not the long-time scaling: the asymptotic decay is universal. In a Hubble time, the Hawking temperature falls well below the electron mass.

\subsection{Relation to earlier work}

These results dovetail with the qualitative analysis of the near-extremal regime given in Ref.~\cite{Lehmann:2019jcap}, which emphasized that RN black holes with $Q \sim M$ undergo substantial modifications to their evaporation dynamics. The vanishing of the surface gravity at extremality forces the Hawking temperature to zero, consistent with the cosmic censorship conjecture, but does not automatically guarantee stability. In fact, while neutral particle emission is forbidden, charged particle emission may still proceed through the Schwinger mechanism, provided $|q| > m$ for the relevant species. As discussed in Ref.~\cite{Lehmann:2019jcap}, this implies that exactly extremal states may still discharge, a perspective that connects to the weak gravity conjecture and its motivation in avoiding an infinite tower of stable charged remnants.

\smallskip\noindent
Moreover, the third law of black hole thermodynamics forbids reaching exact extremality in finite time, so any stalling of the evaporation process can only be approximate. Numerical studies of near-extremal evaporation for large black holes \cite{Hiscock:1990ex} showed that evaporation slows significantly but never halts, a conclusion echoed by our explicit time-domain calculations at the Planck scale.  The implication is that, in the absence of new quantum-gravity physics, charged Planck-scale black holes evolve asymptotically toward extremality, producing charged relics that remain frozen with their residual electric charge. The relic population thus naturally retains a finite charged fraction, consistent with estimates based on stochastic charge fluctuations near the Planck scale.

\smallskip\noindent
For macroscopic Reissner–Nordström black holes, the combined Hawking–Schwinger evolution reduces $Q/M$ monotonically.  
Near the Planck scale, however, the Hawking temperature and the Schwinger rate both vanish as $T_{\rm H}\to0$ and $E_h\to E_{\rm crit}$, producing a quasi-static near-extremal configuration.  
This limiting behaviour corresponds to the freeze-out region described numerically by Hiscock and Weems and analytically by MacGibbon, but here extended to include the explicit environmental and dark-sector effects discussed in the main text.

\bibliographystyle{JHEP}
\bibliography{planck_relic_discharge}

\providecommand{\href}[2]{#2}\begingroup\raggedright\begin{thebibliography}{10}

\bibitem{Gibbons:1975kk}
G.W.~Gibbons, \emph{Vacuum polarization and the spontaneous loss of charge by black holes}, \href{https://doi.org/10.1007/BF01609829}{\emph{Commun. Math. Phys.} {\bfseries 44} (1975) 245}.

\bibitem{MacGibbon:1987}
J.H.~MacGibbon, \emph{{Can Planck-mass relics of evaporating black holes be stable and charged?}}, \href{https://doi.org/10.1038/329308a0}{\emph{Nature} {\bfseries 329} (1987) 308}.

\bibitem{MacGibbon:1988}
J.H.~MacGibbon and B.J.~Carr, \emph{{The charge and baryon-number of evaporating black holes}}, \href{https://doi.org/10.1086/169909}{\emph{Astrophys. J.} {\bfseries 371} (1991) 447} [\href{https://arxiv.org/abs/astro-ph/9204004}{{\ttfamily astro-ph/9204004}}].

\bibitem{Hoyle:1939}
F.~Hoyle and R.A.~Lyttleton, \emph{The effect of interstellar matter on climatic variation}, \href{https://doi.org/10.1017/S0305004100021150}{\emph{Proc. Camb. Phil. Soc.} {\bfseries 35} (1939) 405}.

\bibitem{Bondi:1952}
H.~Bondi, \emph{On spherically symmetrical accretion}, \href{https://doi.org/10.1093/mnras/112.2.195}{\emph{Mon. Not. R. Astron. Soc.} {\bfseries 112} (1952) 195}.

\bibitem{Schwinger:1951nm}
J.~Schwinger, \emph{On gauge invariance and vacuum polarization}, \href{https://doi.org/10.1103/PhysRev.82.664}{\emph{Phys. Rev.} {\bfseries 82} (1951) 664}.

\bibitem{Hiscock:1990}
W.A.~Hiscock and L.D.~Weems, \emph{{Evolution of Charged Evaporating Black Holes}}, \href{https://doi.org/10.1103/PhysRevD.41.1142}{\emph{Phys. Rev. D} {\bfseries 41} (1990) 1142}.

\bibitem{Ewasiuk:2025dwn}
C.~Ewasiuk and S.~Profumo, \emph{{Evolution of Mass, Charge, and Angular Momentum in Black Hole Evaporation: a Comparative Analysis}},  \href{https://arxiv.org/abs/2505.04812}{{\ttfamily 2505.04812}}.

\bibitem{NIST:DLMF}
F.W.J.~Olver, D.W.~Lozier, R.F.~Boisvert and C.W.e.~Clark, ``Nist digital library of mathematical functions.'' \url{https://dlmf.nist.gov/}, 2024.

\bibitem{Lehmann:2019jcap}
B.V.~Lehmann, C.~Johnson, S.~Profumo and T.~Schwemberger, \emph{Direct detection of primordial black hole relics as dark matter}, \href{https://doi.org/10.1088/1475-7516/2019/10/046}{\emph{JCAP} {\bfseries 10} (2019) 046} [\href{https://arxiv.org/abs/1906.06348}{{\ttfamily 1906.06348}}].

\bibitem{Unruh:1976fm}
W.G.~Unruh, \emph{{Absorption Cross-Section of Small Black Holes}}, \href{https://doi.org/10.1103/PhysRevD.14.3251}{\emph{Phys. Rev. D} {\bfseries 14} (1976) 3251}.

\bibitem{Giffin:2021kgb}
P.~Giffin, J.~Lloyd, S.D.~McDermott and S.~Profumo, \emph{{Neutron star quantum death by small black holes}}, \href{https://doi.org/10.1103/PhysRevD.105.123030}{\emph{Phys. Rev. D} {\bfseries 105} (2022) 123030} [\href{https://arxiv.org/abs/2105.06504}{{\ttfamily 2105.06504}}].

\bibitem{Heisenberg:1936bi}
W.~Heisenberg and H.~Euler, \emph{Consequences of dirac’s theory of the positron}, \href{https://doi.org/10.1007/BF01343663}{\emph{Z. Phys.} {\bfseries 98} (1936) 714}.

\bibitem{Carter:1974}
B.~Carter, \emph{Charge and particle conservation in black-hole decay}, \href{https://doi.org/10.1103/PhysRevLett.33.558}{\emph{Phys. Rev. Lett.} {\bfseries 33} (1974) 558}.

\bibitem{Birrell:1982ix}
N.D.~Birrell and P.C.W.~Davies, \emph{Quantum Fields in Curved Space}, Cambridge University Press (1982), \href{https://doi.org/10.1017/CBO9780511622632}{10.1017/CBO9780511622632}.

\bibitem{Wald:2001wrr}
R.M.~Wald, \emph{The thermodynamics of black holes}, \href{https://doi.org/10.12942/lrr-2001-6}{\emph{Living Rev. Relativ.} {\bfseries 4} (2001) 6}.

\bibitem{Heusler:1996}
M.~Heusler, \emph{Black Hole Uniqueness Theorems}, Cambridge University Press (1996), \href{https://doi.org/10.1017/CBO9780511661396}{10.1017/CBO9780511661396}.

\bibitem{Born:1934gh}
M.~Born and L.~Infeld, \emph{Foundations of the new field theory}, \href{https://doi.org/10.1098/rspa.1934.0059}{\emph{Proc. Roy. Soc. Lond. A} {\bfseries 144} (1934) 425}.

\bibitem{Adler:1971wn}
S.L.~Adler, \emph{Photon splitting and photon dispersion in a strong magnetic field}, \href{https://doi.org/10.1016/0003-4916(71)90154-0}{\emph{Annals Phys.} {\bfseries 67} (1971) 599}.

\bibitem{Banks:2010zn}
T.~Banks and N.~Seiberg, \emph{Symmetries and strings in field theory and gravity}, \href{https://doi.org/10.1103/PhysRevD.83.084019}{\emph{Phys. Rev. D} {\bfseries 83} (2011) 084019} [\href{https://arxiv.org/abs/1011.5120}{{\ttfamily 1011.5120}}].

\bibitem{Harlow:2018tng}
D.~Harlow and H.~Ooguri, \emph{Constraints on symmetries from holography}, \href{https://doi.org/10.1103/PhysRevLett.122.191601}{\emph{Phys. Rev. Lett.} {\bfseries 122} (2019) 191601} [\href{https://arxiv.org/abs/1810.05337}{{\ttfamily 1810.05337}}].

\bibitem{Harlow:2018jwu}
D.~Harlow and H.~Ooguri, \emph{Symmetries in quantum field theory and quantum gravity}, \href{https://doi.org/10.1007/s00220-021-04040-y}{\emph{Commun. Math. Phys.} {\bfseries 383} (2021) 1669} [\href{https://arxiv.org/abs/1810.05338}{{\ttfamily 1810.05338}}].

\bibitem{ArkaniHamed:2006dz}
N.~Arkani-Hamed, L.~Motl, A.~Nicolis and C.~Vafa, \emph{The string landscape, black holes and gravity as the weakest force}, \href{https://doi.org/10.1088/1126-6708/2007/06/060}{\emph{JHEP} {\bfseries 06} (2007) 060} [\href{https://arxiv.org/abs/hep-th/0601001}{{\ttfamily hep-th/0601001}}].

\bibitem{Heidenreich:2016aqi}
B.~Heidenreich, M.~Reece and T.~Rudelius, \emph{{Evidence for a sublattice weak gravity conjecture}}, \href{https://doi.org/10.1007/JHEP08(2017)025}{\emph{JHEP} {\bfseries 08} (2017) 025} [\href{https://arxiv.org/abs/1606.08437}{{\ttfamily 1606.08437}}].

\bibitem{Okun:1982xi}
L.B.~Okun, \emph{Limits of electrodynamics: paraphotons?}, {\emph{Sov. Phys. JETP} {\bfseries 56} (1982) 502}.

\bibitem{Galison:1983pa}
P.~Galison and A.~Manohar, \emph{Two z's or not two z's?}, \href{https://doi.org/10.1016/0370-2693(84)91161-4}{\emph{Phys. Lett. B} {\bfseries 136} (1984) 279}.

\bibitem{Jaeckel:2010ni}
J.~Jaeckel and A.~Ringwald, \emph{The low-energy frontier of particle physics}, \href{https://doi.org/10.1146/annurev.nucl.012809.104433}{\emph{Ann. Rev. Nucl. Part. Sci.} {\bfseries 60} (2010) 405} [\href{https://arxiv.org/abs/1002.0329}{{\ttfamily 1002.0329}}].

\bibitem{Fabbrichesi:2020wbt}
M.~Fabbrichesi, E.~Gabrielli and G.~Lanfranchi, \emph{The Dark Photon}, SpringerBriefs in Physics, Springer (2020), \href{https://doi.org/10.1007/978-3-030-62519-1}{10.1007/978-3-030-62519-1}, [\href{https://arxiv.org/abs/2005.01515}{{\ttfamily 2005.01515}}].

\bibitem{CyrRacine:2012fz}
F.-Y.~Cyr-Racine and K.~Sigurdson, \emph{The cosmology of atomic dark matter}, \href{https://doi.org/10.1103/PhysRevD.87.103515}{\emph{Phys. Rev. D} {\bfseries 87} (2013) 103515} [\href{https://arxiv.org/abs/1209.5752}{{\ttfamily 1209.5752}}].

\bibitem{Feng:2009mn}
J.L.~Feng, M.~Kaplinghat, H.~Tu and H.-B.~Yu, \emph{Hidden charged dark matter}, \href{https://doi.org/10.1088/1475-7516/2009/07/004}{\emph{JCAP} {\bfseries 07} (2009) 004} [\href{https://arxiv.org/abs/0905.3039}{{\ttfamily 0905.3039}}].

\bibitem{Holdom:1985ag}
B.~Holdom, \emph{Two u(1)'s and epsilon charge shifts}, \href{https://doi.org/10.1016/0370-2693(86)91377-8}{\emph{Phys. Lett. B} {\bfseries 166} (1986) 196}.

\bibitem{Coleman:1991QuantumHair}
S.R.~Coleman, J.~Preskill and F.~Wilczek, \emph{{Quantum hair on black holes}}, \href{https://doi.org/10.1016/0550-3213(92)90008-Y}{\emph{Nucl. Phys. B} {\bfseries 378} (1992) 175} [\href{https://arxiv.org/abs/hep-th/9201059}{{\ttfamily hep-th/9201059}}].

\bibitem{Davidson:2000hf}
S.~Davidson, S.~Hannestad and G.~Raffelt, \emph{Updated bounds on milli-charged particles}, \href{https://doi.org/10.1088/1126-6708/2000/05/003}{\emph{JHEP} {\bfseries 05} (2000) 003} [\href{https://arxiv.org/abs/hep-ph/0001179}{{\ttfamily hep-ph/0001179}}].

\bibitem{Vogel:2013raa}
H.~Vogel and J.~Redondo, \emph{Dark radiation constraints on minicharged particles in models with a hidden photon}, \href{https://doi.org/10.1088/1475-7516/2014/02/029}{\emph{JCAP} {\bfseries 02} (2014) 029} [\href{https://arxiv.org/abs/1311.2600}{{\ttfamily 1311.2600}}].

\bibitem{Hiscock:1990ex}
W.A.~Hiscock and L.D.~Weems, \emph{Evolution of charged evaporating black holes}, \href{https://doi.org/10.1103/PhysRevD.41.1142}{\emph{Phys. Rev. D} {\bfseries 41} (1990) 1142}.

\end{thebibliography}\endgroup
\end{document}